\documentclass[10pt,english]{article}
\usepackage{mathptmx}
\usepackage[T1]{fontenc}
\usepackage[latin9]{inputenc}
\usepackage{geometry}
\usepackage{color}
\usepackage{babel}
\usepackage{float}
\usepackage{amsmath}
\usepackage{amsthm}
\usepackage{graphicx}
\usepackage{esint}
\usepackage{amsfonts}
\usepackage{amssymb}
\usepackage{lineno}
\usepackage{stackrel}
\usepackage{units}
\usepackage{mathrsfs}
\usepackage{setspace}
\usepackage[authoryear]{natbib}
\usepackage{soul}
\usepackage[unicode=true,pdfusetitle,
 bookmarks=true,bookmarksnumbered=true,bookmarksopen=true,bookmarksopenlevel=13,
 breaklinks=false,pdfborder={0 0 0},pdfborderstyle={},backref=false,colorlinks=true]
 {hyperref}

\usepackage{url} 
\usepackage[export]{adjustbox}[2011/08/13]

\makeatletter
\newcommand{\lyxaddress}[1]{
	\par {\raggedright #1
	\vspace{1.4em}
	\noindent\par}
}

\@ifundefined{date}{}{\date{}}
\usepackage{color}
\usepackage[usenames,dvipsnames,svgnames,table]{xcolor}
\usepackage{multicol}
\definecolor{burgundy}{rgb}{0.5, 0.0, 0.13}
\definecolor{airforceblue}{rgb}{0.36, 0.54, 0.66}
\hypersetup{urlcolor=airforceblue}
\hypersetup{citecolor=burgundy}

\setcitestyle{notesep={, },super,open={},close={},comma,aysep={,},yysep={,}}  
 \makeatletter
   \renewcommand\@biblabel[1]{#1.}
 \makeatother

\makeatother

\begin{document}
\title{The westward drift of Jupiter's polar cyclones explained by a center-of-mass
approach}
\author{\href{https://orcid.org/0000-0002-3645-0383}{Nimrod Gavriel}$^{1\star}$
and \href{https://orcid.org/0000-0003-4089-0020}{Yohai Kaspi}$^{1}$}
\maketitle

\lyxaddress{\begin{center}
\textit{$^{1}$Department of Earth and Planetary Sciences, Weizmann
Institute of Science, Rehovot, Israel}\\
\textit{$^{\star}$\href{mailto:nimrod.gavriel@weizmann.ac.il}{nimrod.gavriel@weizmann.ac.il}}
\par\end{center}}

\lyxaddress{\begin{center}
Preprint September 10, 2023\textit{}\\
\textit{Geophys. Res. Lett.}. 50, 19, (2023). DOI:\href{https://doi.org/10.1029/2023GL103635}{10.1029/2023GL103635}
\\
(Received Mars 10, 2023; Revised July 18, 2023; Accepted August
29, 2023) 
\par\end{center}}
\begin{abstract}
The first orbits around Jupiter of the Juno spacecraft in 2016
revealed a symmetric structure of multiple cyclones that remained
stable over the next five years. Trajectories of individual cyclones
indicated a consistent westward circumpolar motion around both poles. In this paper,
we propose an explanation for this tendency using the concept of beta-drift
and a "center-of-mass" approach. We suggest that the motion of these
cyclones as a group can be represented by an equivalent sole cyclone,
which is continuously pushed by beta-drift poleward and westward,
embodying the westward motion of the individual cyclones. We support
our hypothesis with 2D model simulations and observational evidence,
demonstrating this mechanism for the westward drift. This study joins
consistently with previous studies that revealed how aspects of these
cyclones result from vorticity-gradient forces, shedding light on
the physical nature of Jupiter's polar cyclones.
\end{abstract}


\twocolumn 

\subsection*{Plain Language Summary}
The Juno spacecraft arrived at Jupiter in 2016, revealing a
unique atmospheric phenomenon. Each of the poles of Jupiter is inhabited
by a symmetrically structured group of cyclones, where a ring of cyclones
surrounds one cyclone close to the pole. The collective observations
of these cyclones over five years show that although they are relatively
stable, they generally drift in the westward direction a few degrees
per year. Here, we investigate the mechanism driving this drift by
examining the cyclones as a group. This \textquotedbl center-of-mass\textquotedbl{}
approach masks the interactions between the cyclones and only considers
trends that happen simultaneously on all cyclones. Using model simulations,
we show that the motion of a group's \textquotedbl center of mass\textquotedbl{} can be captured
by a sole equivalent cyclone, pushed poleward and westward by the
\textquotedbl beta-drift\textquotedbl{} effect, which is known to
contribute to the motion of tropical cyclones on Earth. This westward
force on the group as a whole is thus suggested as the driver of the
observed westward drift. We conclude by presenting observational evidence
supporting this hypothesis. 
\subsection*{Key Points:}
\begin{itemize}
\item The mean westward circumpolar motion of Jupiter's northern and southern polar cyclones is analyzed and explained by the $\beta$-drift effect
\item Simulations show that the "center of mass" of a group of cyclones behaves like one equivalent cyclone, moving poleward-westward under $\beta$-drift
\item This center-of-mass approach is applied to the Juno data, implying that the cyclones' collective $\beta$-drift drives their observed westward drift 
\end{itemize}

\section{Introduction}
\begin{figure*}[ht!]
\includegraphics[width=1\textwidth,center]{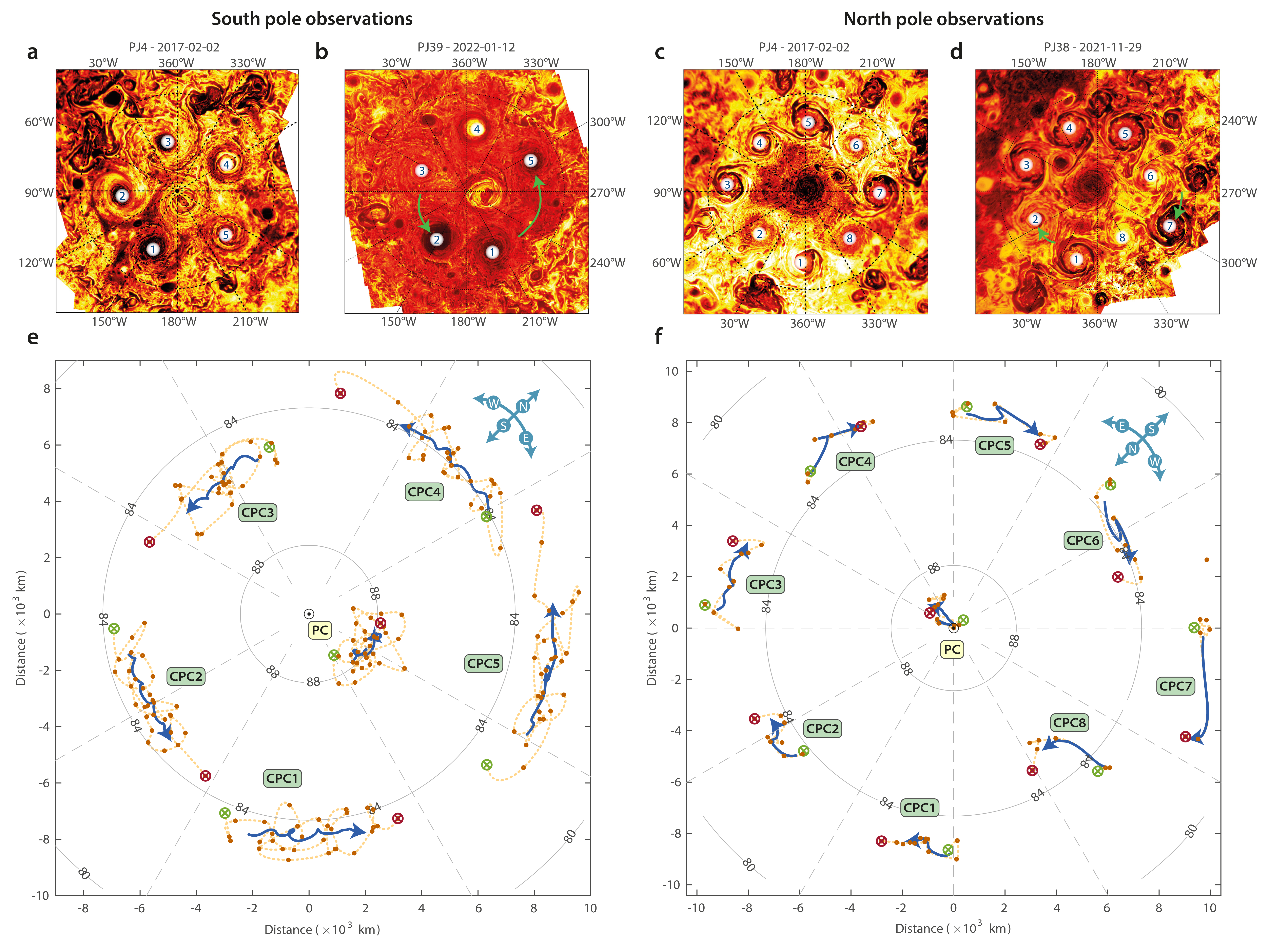}
\caption{Observations of a westward drift in the north and south poles of Jupiter.
(a-d) Infrared images of Jupiter's poles taken by the Jovian Infrared
Auroral Mapper (JIRAM) instrument onboard Juno. These panels were
adapted from Ref.\cite{mura2022five_years}. (a and b) Images from the
south pole in February 2017 (panel a), and in January 2022 (panel
b). (c and d) Images from the north pole in February 2017 (panel c),
and in November 2021 (panel d). (e and f) Trajectories of the cyclones
in the south (panel e) and north (panel f) poles, as observed along
5 years. The orange dots represent the observed locations during Juno's
polar approaches \cite{mura2021oscillations,mura2022five_years}.
The dashed yellow curves represent an interpolation of the trajectories between
observations. The green and red crosses represent the observed locations
in February 2017 and five years later, respectively. The blue arrows
represent a moving average of the interpolated trajectories, showing
the overall westward trend experienced by the individual cyclones.
The gray circles denote the $88^{\circ}, 84^{\circ}$ and $80^{\circ}$
latitude lines. The gray dashed lines represent increments of $30^{\circ}$
longitude. The point (0,0) represents the pole.
}
\label{fig: cart_trajectories}
\end{figure*}
The polar cyclones of Jupiter, revealed in 2017 by the Juno mission
\cite{bolton2017jupiter,orton2017first,adriani2018}, are comprised
of a single polar cyclone (PC) situated near each pole, and multiple
circumpolar cyclones (CPCs), 5 at the south pole and 8 at the north
pole. The CPCs surround each PC in a ring at approximately latitude
$\pm84^{\circ}$ (Fig.~\ref{fig: cart_trajectories}).
While the cyclones were observed to retain a stable structure during
the five years of measurements, slow motions about their mean positions,
and a general westward trend were reported for both poles (Fig.~\ref{fig: cart_trajectories};
References\cite{adriani2020two,tabataba2020long,mura2021oscillations,mura2022five_years}).
The individual cyclones maintain maximum winds of up to $\sim100$
ms$^{-1}$ at $\sim1,000$ km from their cores, and extend to diameters
about $5,000$ km \cite{grassi2018first,adriani2020two}. While a
stable configuration of CPCs were never observed before Juno's polar
approaches, a polar cyclone has been persistently observed at each
pole of Saturn \cite{baines2009saturn}. The formation of a polar
cyclone on a gas giant is well understood and revolves around the
concept of ``beta-drift'', which drives cyclonic vortices poleward
\cite{rossby1948displacements,adem1956series,fiorino1989some,smith1990analytical},
where they converge and maintain a polar cyclone \cite{scott2011polar,oneill2015polar,oneill2016,brueshaber2019dynamical,hyder2022exploring}.
This $\beta$-drift mechanism will be further explained in the following
sections.

Given this polar attraction of cyclones, an additional mechanism is
needed for explaining the existence of the stable Jovian CPCs, so
that they would not merge at the poles. The gradient of planetary
vorticity, $\beta$, is the driver of $\beta$-drift. It was shown
that if, in addition to $\beta$, the vorticity gradient of the PCs
is accounted for when considering $\beta$-drift, then the Jovian
CPCs are located at a latitude (approximately $\pm84^{\circ}$) where
the net vorticity gradient vanishes, and thus the CPCs are in a stable
equilibrium \cite{gavriel2021number}. On Saturn, which is dynamically
similar to Jupiter \cite{showman2018global,galanti2019,kaspi2020comparison},
such equilibrium is currently unattainable, inhibiting stable CPCs
\cite{gavriel2021number}. For the PCs to have a vorticity gradient
with the right direction to oppose $\beta$, they need to posses an
anticyclonic ``shielding'' around them, as was shown using Shallow-Water
(SW) simulations \cite{li2020}. 

\subsection{Observed motion of Jupiter's polar cyclones}

In Fig.~\ref{fig: cart_trajectories}e-f, the reported
locations of Jupiter's polar cyclones \cite{mura2022five_years} during the period 2017-2022
are presented. As direct observations of the north polar cyclones
were much less frequent than those of the south, the trajectories of the
north cyclones are much less constrained. Focusing on the better determined
trajectories at the south pole (Fig.~\ref{fig: cart_trajectories}e),
two trends become apparent. One motion trend is a mostly circular
motion with a period of $\sim12$ months and a radius of $\sim400$
km \cite{gavriel2022oscillatory}. This motion was shown to follow
a strong correlation between the instantaneous acceleration of each
cyclone to the estimated total vorticity gradient under which it moves,
suggesting that this motion is driven by the generalized $\beta$-drift
mechanism \cite{gavriel2022oscillatory}. The second motion trend
is an overall westward tendency of each cyclone (Blue arrows in Fig.~\ref{fig: cart_trajectories}e-f),
averaging to about $3^{\circ}$ longitude per year at the north pole
and $7^{\circ}$ at the south \cite{mura2022five_years}. In this
study, we provide an explanation for this westward trend by considering
the generalized $\beta$-drift of all the polar cyclones at large.
In the following section we explain how one cyclone in a $\beta$-plane
would feel an acceleration poleward and westward due to $\beta$-drift.
Then we suggest a ``center-of-mass'' (CM) approach where the aggregate
motion of a group of cyclones can be estimated by the motion of one
``equivalent cyclone'' following the group's CM, which allows
filtering the noisy interactions between the cyclones. Thus, the equivalent
cyclone, orbiting the pole due to the poleward component of $\beta$-drift
would also precess westward due to the westward component of $\beta$-drift,
projecting on the westward motion of the individual cyclones. We conclude
our analysis with observational support for this hypothesis.

\section{The evolution of poleward-westward acceleration of barotropic cyclones
with a background vorticity gradient}

\begin{figure*}[ht!]
\includegraphics[width=1\textwidth,center]{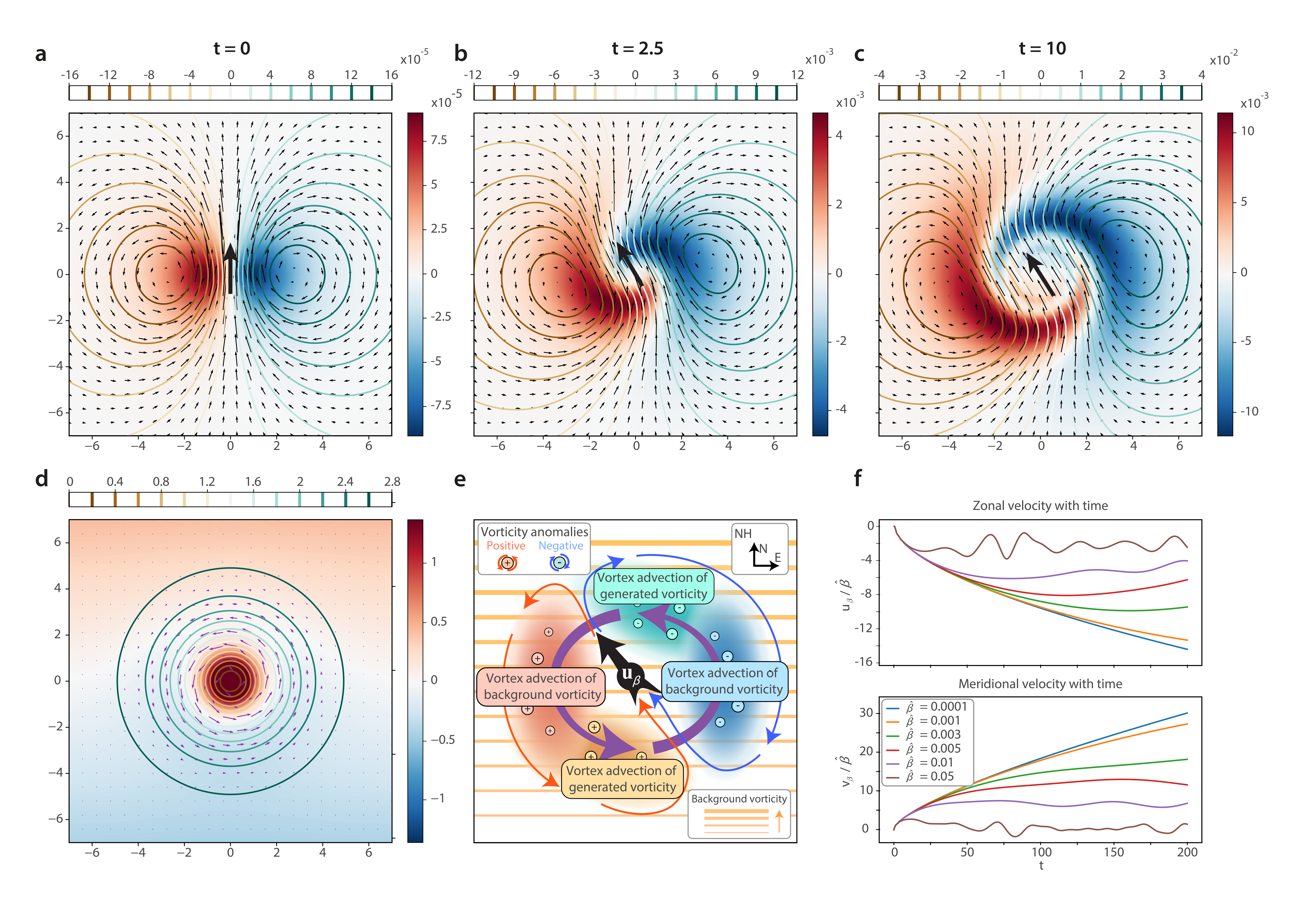}
\caption{The mechanism for the westward component of $\beta$-drift. (a-c) Three
snapshots from a simulation showing the development of a $\beta$-drift
secondary circulation. The red-blue contours represent the generated
vorticity. The brown-teal contours are streamlines. Black arrows represent
the direction and relative magnitude of the generated velocity field,
where the big central arrow represents the instantaneous $\beta$-drift
advective velocity, at which the cyclone is traveling. All the numbers
are unit-less, normalized by the respective cyclone characteristic
parameters. (d) The same as\textbf{ }panels a-c, but for the primary
cyclone circulation forcing the simulation. The north-south color
gradient represents the gradient of background vorticity (for illustrative
purposes, this gradient is enhanced relative to the values used in
the simulation). A video of the simulation is available in SI movie
1. (e) An illustration of the two main terms ($\times2$ for each
side) that determine $\beta$-drift, according to Eq.~\ref{eq: Secondary circ}.
The purple arrow represents the cyclone's tangential velocity (\textbf{$\mathbf{u}_{{\rm v}}$}).
The orange lines represent the background vorticity, where thicker
lines represent larger magnitudes. The red-orange (blue-turquoise)
clouds with plus (minus) signs represent generation of positive (negative)
vorticity. The red (blue) arrows represent the generated cyclonic
(anti-cyclonic) secondary circulation. The black arrow is the resulting
$\beta$-drift velocity. While this illustration is for the northern
hemisphere, the same principles work for the southern hemisphere,
where a poleward-westward drift is generated as well (see the equivalent
Fig.~S1). (f) Time evolution of the simulated $\beta$-drift velocity
at the center of the cyclone, divided by $\hat{\beta}$ (black arrow
in the center of panels a-c). Different colors represent different
$\hat{\beta}$ values, showing that for a significant period of the
acceleration, the velocities are proportional to $\hat{\beta}$.}
\label{fig:Theory fig}
\end{figure*}
To understand how $\beta$-drift can cause the observed westward motion
of Jupiter's polar cyclones, we start by reexamining the ideal case
of a 2D (barotropic) cyclone under a linear change in background vorticity ($\beta$-plane).
This scenario has been studied extensively in the past in the context
of tropical cyclones on Earth, establishing the concept of $\beta$-drift,
which is the generation of a dipole of vorticity (termed $\beta$-gyres)
propelling cyclones in a tilted direction relative to the gradient
of background vorticity \cite{rossby1948displacements,adem1956series,fiorino1989some,smith1990analytical,smith1990numerical,smith1993vortex}.
Here, we illustrate the generation of the $\beta$-gyres by simulating,
using the Dedalus solver \cite{burns2020dedalus}, the secondary vorticity
tendency according to the partitioning suggested by Ref.\cite{smith1990numerical}.
The scaled equation we solve is, at leading order, 
\begin{linenomath*}
\begin{equation}
\frac{\partial\xi_{{\rm g}}}{\partial t}=-\hat{\beta}v_{{\rm v}}-\mathbf{u}_{{\rm v}}\cdot\mathbf{\nabla}\xi_{{\rm g}},\label{eq: Secondary circ}
\end{equation}
\end{linenomath*}
where $\xi_{{\rm g}}$ is the secondary (generated) relative vorticity, 
and $\mathbf{\mathbf{u}_{{\rm v}}}$ is a prescribed idealized vortex
velocity vector with components $\left(u_{{\rm v}},v_{{\rm v}}\right)$
in the $\left(x,y\right)$ directions, in a grid moving such that
the vortex is always at the center (see SI for the full equations
and other model details; the used scale factors are supplied in Eq.~S6). The first term on the RHS, the advection
of background potential vorticity by the vortex (originating from
the curl of the Coriolis force), is proportional to the unit-less
parameter $\hat{\beta}=\nicefrac{\Delta f}{\omega}$, which is the
ratio between the change in background vorticity across the core
of the vortex ($\Delta f=\beta R$, where $R$ is the radius of maximum
velocity in the vortex), and the rotation frequency scale of the vortex ($\omega=\nicefrac{V}{R}$,
where $V$ is the maximum velocity in the vortex). Thus, this parameter
encapsulates how steep is the vorticity gradient across the cyclone,
which in turn would determine the resulting magnitudes of $\xi_{{\rm g}}$.
In addition, the second term on the RHS of Eq.~\ref{eq: Secondary circ}
represents the subsequent advection of generated $\xi_{{\rm g}}$
by the primary circulation of the vortex. The full solved equation also contains the nonlinear terms, a numerical viscosity term and a sponge term which helps avoid effects from the double periodic boundaries.

In Fig.~\ref{fig:Theory fig}a-c,\textbf{ }three simulation snapshots
are plotted for $\xi_{{\rm g}}$, resulting from the primary vortex
circulation in a $\beta$-plane, as shown in Fig.~\ref{fig:Theory fig}d.
As the simulation begins from rest $\left(\xi_{{\rm g}}=0\right)$,
the only term generating vorticity immediately after $t=0$ (Fig.~\ref{fig:Theory fig}a)
is the $-\hat{\beta}v_{{\rm v}}$ term, which changes according to
the meridional velocity of the primary circulation. As such, near
$t=0$, we have a dipole oriented in the $x$ direction, creating
an initial acceleration northward (towards the gradient of background
vorticity). This $\xi_{{\rm g}}$ generation is illustrated\textbf{
}as the red and blue regions in Fig.~\ref{fig:Theory fig}e, representing
the advective flux of background vorticity by $\mathbf{u}_{{\rm v}}$.
Then, this generated $\xi_{{\rm g}}$ dipole is twisted by $\mathbf{u}_{{\rm v}}$
(Fig.~\ref{fig:Theory fig}b-c and orange and turquoise clouds in
Fig.~\ref{fig:Theory fig}e), accelerating the vortex in the westward
direction. The direction of the $\beta$-drift velocity vector ($\mathbf{u}_{{\rm \beta}}$)
is determined by the diagonal line where the terms $-\hat{\beta}v_{{\rm v}}$
and $-\mathbf{u}_{{\rm v}}\cdot\mathbf{\nabla}\xi_{{\rm g}}$ cancel
each other. This direction is poleward-westward for cyclones with
background vorticity increasing poleward, but would flip to equatorward-eastward
if the background vorticity increases equatorward (Fig.~S1), such as in the case when the Jovian CPCs get too close to the shielded PCs \cite{gavriel2021number, gavriel2022oscillatory}. 

An account of the acceleration of $\beta$-drift with time can be seen
in Fig.~\ref{fig:Theory fig}f, showing how the secondary velocities
at the center of the cyclone ($u_{\beta}$ and $v_{\beta}$ in the
$x,y$ directions, respectively) are proportional to $\hat{\beta}$
for the majority of the acceleration phase. If the velocities were
not scaled by $\hat{\beta}$, the different curves would have magnitudes
proportional to $\hat{\beta}$, but this scaling fuses the curves
together. The curves eventually separate due to the accumulation of
non-linearities. This separation happens sooner for simulations with
larger $\hat{\beta}$. The direction of $\mathbf{u}_{\beta}$ can
be calculated from the two panels of Fig.~\ref{fig:Theory fig}f
as $\tan^{-1}\left(\nicefrac{v_{\beta}}{u_{\beta}}\right)$. The time
is scaled by $\nicefrac{R}{V}$, which is of order $\sim3$ hours
for Jupiter's cyclones; the velocity is scaled by $V$, in addition
to $\hat{\beta}$. It can be seen that the model is unstable for $\hat{\beta}$
values on the order of $5\times10^{-2}$ and larger, which we assume
to be due to the non-linear effects. The $\hat{\beta}$ values of
Jupiter's polar cyclones are between $10^{-4}$ to $10^{-3}$ (Fig.~\ref{fig: CM observations}),
and thus are represented well by the linear solutions. Another thing to note here about the applicability of Fig.~\ref{fig: CM observations}f to the Jovian polar cyclones is that as the cyclones move around, the effective $\beta$ under which they operate changes its magnitude and direction, resulting in a resetting (or partial resetting) of the $\beta$-drift phase seen in Fig.~\ref{fig: CM observations}f and limiting the late-time evolution of the $\beta$-drift.

\section{The center of mass for a group of cyclones and the motion of an equivalent
cyclone}

\begin{figure*}[ht!]
\includegraphics[width=1\textwidth,center]{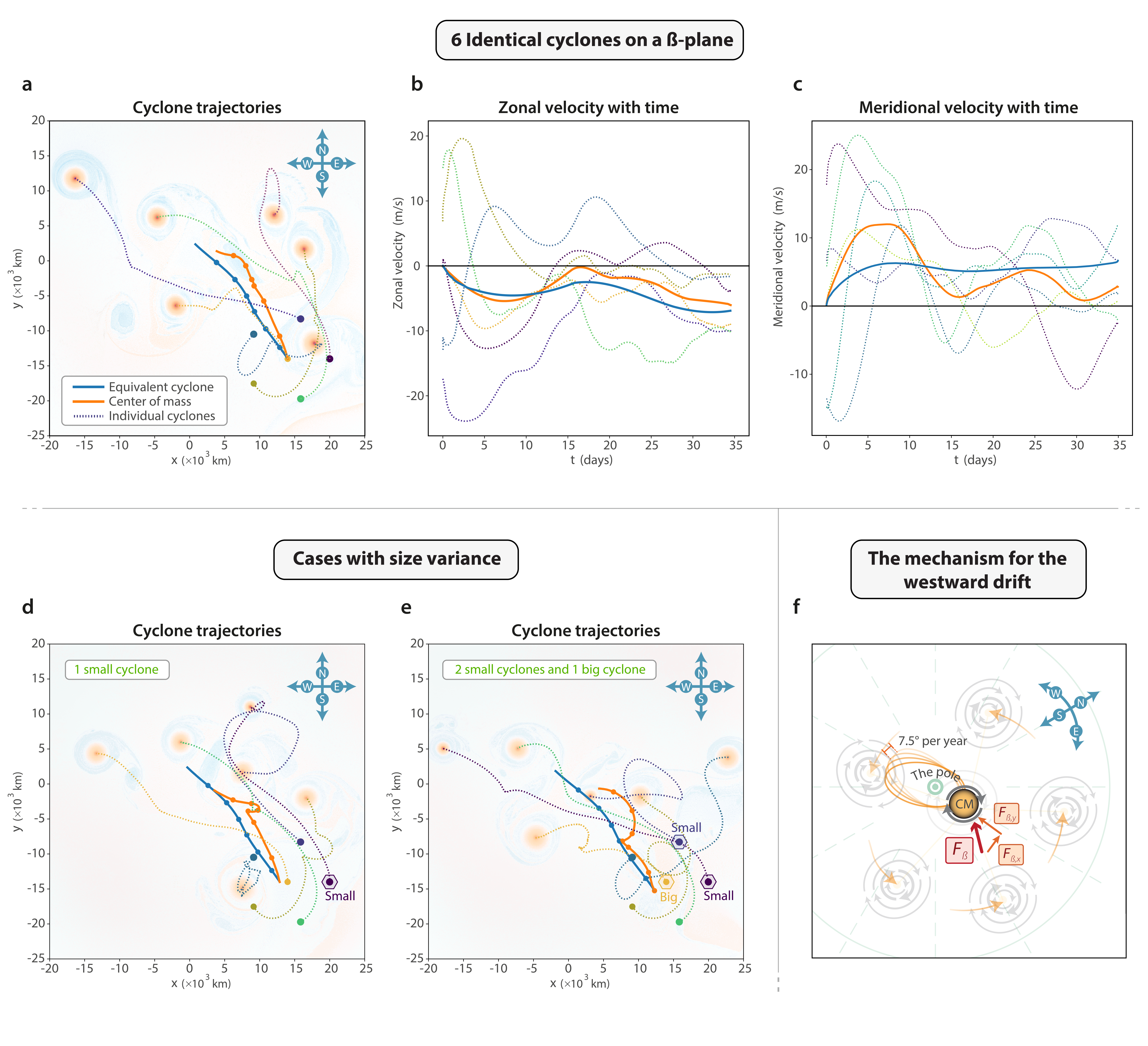}
\caption{A CM approach for a group of cyclones. (a-e) The simulated
motion of 6 cyclones on a $\beta$-plane and of their equivalent cyclone. 
(a) The Cartesian trajectories with time. The colored dots are the
initial locations of 6 identical cyclones. The colored dotted lines are the
trajectories of the individual cyclones. The orange (blue) curve represents
the trajectory of the CM (equivalent cyclone), where the dots represent
constant time intervals. The red-blue contour represent the relative
vorticity map at the final snapshot. The colors are enhanced for illustrative
purposes. (b and c) The same as panel a, but here the abscissa is time and the ordinate
is zonal and meridional velocity, respectively. (d) The same as panel a, but in this case one of the cyclones is smaller 
by $50\%$. (e) Another simulation, similar to panel a, where two cyclones are smaller by $50\%$, and one cyclone
is larger by $20\%$. See SI Movie 2 for a video of the entire simulation of the
6 cyclones and SI Movie 3 for a video of the equivalent cyclone simulation (for the case with identical cyclones).
(f) An illustration of how the poleward-westward
$\beta$-drift acceleration can create a net westward drift in Jupiter's
poles when looking from the perspective of a CM.
The transparent cyclones represent the CPCs. The red arrow is the
$\beta$-drift acceleration, where the orange arrows are the meridional
and zonal components. The orange ellipses illustrate the motion of
the CM around the pole, where the ellipse precesses westward due to
the zonal component of the $\beta$-drift.}
\label{fig:Center of mass simulation}
\end{figure*}
In previous numerical studies of tropical cyclones, it was shown that
two identical cyclones on an $f$-plane may orbit each other due to
their mutual interaction, but once a poleward background vorticity gradient ($\beta$) is introduced, both cyclones
gain a poleward-westward motion, while their motion relative to one
another stays the same as in the $f$-plane scenario \cite{chan1995interaction}.
Thus, it can be inferred that an equivalent cyclone positioned in
the CM of the cyclones, moves according
to $\beta$-drift while absorbing the mutual interactions. To test
this idea further, we use Dedalus to solve the barotropic
vorticity equation, this time without the partitioning to primary
and secondary circulations as in Fig.~\ref{fig:Theory fig}, so that
\begin{linenomath*}
\begin{equation}
\frac{D\xi}{Dt}=-\beta v,\label{eq: abs vorticity equation}
\end{equation}
\end{linenomath*}
where $\nicefrac{D}{Dt}=\left(\nicefrac{\partial}{\partial t}+\mathbf{u}\cdot\nabla\right)$
is the material derivative, $\mathbf{u}$ is the velocity vector (with
components $u$ and $v$ in the $x$ and $y$ directions, respectively)
and $\xi=\nabla\times\mathbf{u}$ is relative vorticity. Similar to the model in the previous section, here also a numerical viscosity term and a trap term (analogous to the one used in Ref.\cite{siegelman2022polar}), are added to the model.

Eq.~\ref{eq: abs vorticity equation} is integrated on a $\beta$-plane from an initial condition of a group of 6 identical cyclones
positioned according to the colored dots in Fig.~\ref{fig:Center of mass simulation}a (see full model specifications in the SI). The dashed curves are
the Cartesian trajectories of the 6 cyclones and the orange curve
represents the motion of their CM, calculated as the arithmetic mean of the instantaneous positions of the cyclones. In addition, another simulation is
presented where only one equivalent cyclone, identical to the cyclones
in the group, begins its motion from the center of the group and follows
almost the exact trajectory of the CM. Thus, the motion is decoupled
to two contributions, one is the north-westward motion of each cyclone
due to $\beta$-drift, taking only $\beta$ into account, and the
other is the mutual interactions between the cyclones, which are absorbed
when following the CM. The slight difference between the trajectories
of the CM and the equivalent cyclone likely owes to non-linearities. To evaluate how should the cyclones be weighted in order
to determine the CM when cyclones are not identical, we derive a weight estimation (see derivation in the SI) 
based on integrated vorticity gradient forces between two cyclones (Fig.~S2), leading to
\begin{linenomath*}
\begin{equation}
{W}_i= \frac{e^{-\frac{L}{R_{i}}} R_{i}^{-4}}{\sum\left(e^{-\frac{L}{R_{j}}} R_{j}^{-4}\right)},\label{eq:cyc_weight}
\end{equation}
\end{linenomath*}
where ${W}$ is relative weight, $i$ (or $j$) is the cyclone's index, $L$ is an average distance between the cyclones and the sum is over all the cyclones in the group.
With this weight estimation, two other test cases are performed (Fig.~\ref{fig:Center of mass simulation}d-e), similar to Fig.~\ref{fig:Center of mass simulation}a-c, but where the cyclones vary in their sizes, leading to similar results.

The implication of this idealized case to Jupiter's polar cyclones
is non-trivial due to, e.g., variations in the cyclones' properties
within the group and the non-linear background vorticity (which behaves like a cosine of latitude). Our hypothesis is thus (Fig.~\ref{fig:Center of mass simulation}f),
that if we look on an equivalent cyclone in the polar case, this cyclone
only feels the acceleration due to $\beta$-drift, with no dependence on the inter-cyclone interactions. This means that while a CPC would
feel alternating poleward-westward and equatorward-eastward accelerations
by $\beta$ and by interactions with the vorticity gradient of the
other cyclones \cite{gavriel2022oscillatory}, on average, the residual
is only due to $\beta$. The poleward component of the $\beta$-drift
on the CM's equivalent cyclone would then maintain an elliptical orbit
around the pole, and the westward component would rotate this orbit
in the zonal direction. It is therefore proposed that the projection
back to the cyclones of this rotation is the mechanism behind the
observed westward drift at Jupiter's poles. In the following section
we use Juno's observations to support this hypothesis.

\section{The center of mass of the polar cyclones in the observations.}

\begin{figure*}[ht!]
\includegraphics[width=1\textwidth,center]{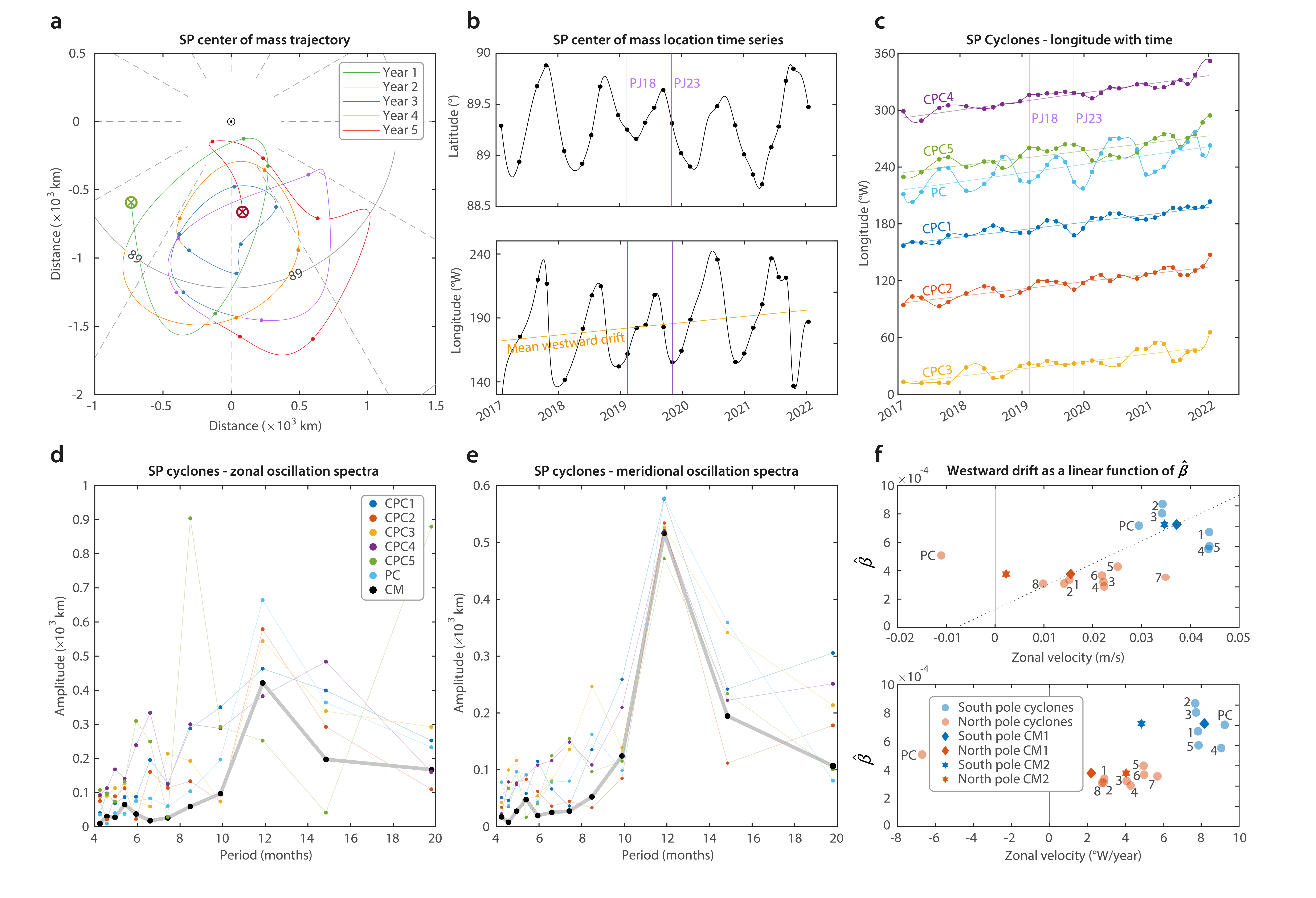}
\caption{Calculations of the CM from the observations. (a) The Cartesian trajectory
of the CM in the south pole of Jupiter according to the cyclone trajectories
presented in Fig.~\ref{fig: cart_trajectories}e (dots
and curves represent real data and interpolation, respectively). The
curve changes color with the passing of every year to illustrate the
1-year oscillatory motion exhibited in the CM's motion. (b) The latitude
(upper panel) and longitude (lower panel) of the south CM with time.
The purple vertical lines represent two incidents of an intruder cyclone
during PJs 18 and 23. The orange line represents the linear fit of
the westward location of the CM with time. (c) The longitude with
time of each of the south polar cyclones. The straight lines represent
linear fits for each cyclone, showing the westward motion. (d and e) The zonal and meridional oscillation
spectra of the south polar cyclones (color) and of the CM (black).
(f) The relation between the measured zonal velocities of the cyclones
and $\hat{\beta}$. Zonal velocities are 5-year averages in metric
units at the upper panel and longitude ($^{\circ}$W) per year in
the lower panel. Red and blue points represent values for the north
and south poles, respectively. The numbers identify the respective CPCs.
The diamond and star shapes represent two different approaches for
calculating the CM's zonal velocities as discussed in the text.}
\label{fig: CM observations}
\end{figure*}
The most straightforward way to calculate the CM is by taking the arithmetic mean of the locations
of all the polar cyclones (the Cartesian projected locations, $\left(x,y\right)$; see Eq.~S1). 
 To take into account the different sizes of the cyclones, which were previously
estimated from Juno's observations \cite{adriani2020two}, we use Eq.~\ref{eq:cyc_weight} for a weighted average of the Cartesian coordinates when determining the CM's trajectory along the period of the observations  (See SI for the calculation procedure and the used values).
As the data from the north pole is too infrequent, we only use the data from
the south pole for Fig.~\ref{fig: CM observations}a-e (Ref.\cite{mura2022five_years},
see Fig.~S3 for the equivalent north-pole figure, based on the
existing observations). It can be seen (Fig.~\ref{fig: CM observations}a)
that the CM's trajectory is indeed very clean, and possesses mostly
the 12-month oscillations exhibited by the individual cyclones (Fig.~\ref{fig: CM observations}d-e,
Ref.\cite{gavriel2022oscillatory}). This suggests that this 12-month
mode of motion is mostly synchronized between the cyclones, and is
therefore largely due to the motion of the CM (i.e., the first mode
of motion where all cyclones oscillate together as a group). 
In both the zonal and meridional cases, short-period modes are suppressed
for the CM (Fig.~\ref{fig: CM observations}d-e). When considering an elliptical orbit with a radial force in the direction of the pole (Fig.~\ref{fig:Center of mass simulation}f), the ellipse is expected to have its semi-major axis in the meridional direction. Consistently, the 12-month amplitude of the CM is larger in the meridional direction than in the zonal (Fig.~\ref{fig: CM observations}d-e).

One conflict between
the hypothesis (Fig.~\ref{fig:Center of mass simulation}f) and the
CM measurement (Fig.~\ref{fig: CM observations}a) is that the measured
CM trajectory does not encompass the pole. This might be due to errors in estimating the weights of the cyclones (e.g., measurement errors, sparsity of data, and approximations in deriving Eq.~\ref{eq:cyc_weight}), which may also change in time during the observation period for the real cyclones. To test the sensitivity of the results (Fig.~\ref{fig: CM observations}) to the prescribed weights and the factor $L$ (Eq.~\ref{eq:cyc_weight}), an analysis of randomly generated cyclone sizes is concluded, showing that the observation-based sizes significantly improve the results when comparing equivalent analyses with random or with identically sized cyclones (Fig.~S4). Also, investigating the role of the parameter $L$ (Fig.~S5), we find that the used value of $6,000$~km, representing an observationally based average distance between the cyclones (see Fig.~\ref{fig: cart_trajectories}e-f), gives the best results in terms of how close is the CM's motion to encompassing the pole.

The time-series of the CM's motion (Fig.~\ref{fig: CM observations}b)
indeed appears to have a clean sinusoidal form, where the inter-cyclone
interactions of the individual cyclones (Fig.~\ref{fig: CM observations}c)
are filtered. The linear fit for the CM's longitude with time (Fig.~\ref{fig: CM observations}b)
has a slope of $8.17^{\circ}$ longitude per year. The analogous linear
fits for the individual cyclones (Fig.~\ref{fig: CM observations}c)
have slopes of $8.23\pm0.71^{\circ}$ longitude per year, suggesting
that indeed the zonal drift of the cyclones is a projection of the
CM's zonal drift (see the lower panel of Fig.~\ref{fig: CM observations}f
for all values, including the north pole). Two reported incidents
where an ``intruder'' vortex appeared in the south polar ring (PJ
18 \& 23, References\cite{adriani2020two,mura2021oscillations}) are marked
(purple vertical lines in Fig.~\ref{fig: CM observations}b-c) as
a possible explanation for the perturbations in the sinusoidal motion
of the CM proximate to these observations. 

Finally, the results from
Figs.~\ref{fig: cart_trajectories}-\ref{fig: CM observations}
are tied together in Fig.~\ref{fig: CM observations}f. Here, the values of $\hat{\beta}$ for the cyclones and for the CM are estimated. As the effective $\beta$ for the individual cyclones is close to zero, representing perturbations near stable equilibrium, the values of $\beta$ (for estimating $\hat{\beta}_i$ of each cyclone $i$) are set equal for all of the cyclones and for the CM at each pole, and are determined by a vector sum of the $\beta$ each cyclone is subject to (see SI). It can be seen (Fig.~\ref{fig: CM observations}f) that indeed $\hat{\beta}$ of the CMs in the north and south poles
are proportional to their respective zonal motion. Here we calculate the CM motion
using two different methods (See SI for the detailed calculations).
For CM1, the velocities are the weighted (according to Eq.~\ref{eq:cyc_weight}) average of the individual
cyclones' zonal velocities. For CM2, the velocities are calculated
by a linear fit of the instantaneous zonal displacements of the CM. Using Fig.~\ref{fig:Theory fig}f with the values of $\hat{\beta}\approx 7\times 10^{-4}$, we get an overestimation in the scale of the zonal velocities of an order of magnitude relative to the south pole values in the upper panel of Fig.~\ref{fig: CM observations}f. Our interpretation of this discrepancy is that the steady state condition may have a form of dissipation which balances the $\beta$-drift acceleration at low values of zonal velocities. Another consideration is a relative vorticity turbulent mixing that may reduce the vorticity gradient of one cyclone under another cyclone \cite{siegelman2022polar} relative to our estimations, which are extrapolated from the profiles at the vicinity of the cyclones. Nevertheless, when using the method CM1 for the metric zonal velocity, which is the more physical form of velocity as related to the $\beta$-drift,
it can be seen that both $\hat{\beta}$ and the CM's zonal velocity are approximately two times larger in the south pole than in the north pole (dashed line in the upper panel of Fig.~\ref{fig: CM observations}f). This proportionality is expected to be true regardless of dissipation or turbulent mixing, and provides support to the given hypothesis for the westward drift in Jupiter's poles.

\section{Discussion}

Since the discovery of Jupiter's polar cyclones, the role of vorticity
gradient forces (or generalized $\beta$-drift), which takes into
account both $\beta$ and the relative vorticities of all neighboring
cyclones, presented a consistent picture. These forces were used to
explain the mean latitude of the cyclones and their number \cite{gavriel2021number},
the oscillatory motion patterns of the cyclones \cite{gavriel2022oscillatory},
and now their mean westward drifts as well. This series of studies,
revealing different aspects of these forces, implies that these polar
cyclones behave, to leading order, like discrete objects, linearly
forced by a ``spring-like'' system driving them all in the poleward-westward
direction while pushing them one from another. There are many uncertainties
in the observations, such as variations between the velocity profile
of each cyclone (and how it changes in time\cite{scarica2022stability})
and the low resolution of the cyclones' tracks \cite{mura2022five_years}, especially in the north pole,
in addition to complexities in the conversion of the CM concept from
Cartesian to concentric polar coordinates and others. 

However, we see that the CM of the cyclones has an organized
motion (Fig.~\ref{fig: CM observations}), filtering the inter-cyclone
exchanges as in the idealized case (Fig,~\ref{fig:Center of mass simulation}a-e),
and that the relative westward velocities of the CMs in the north
and south poles, representatives of ``equivalent'' cyclones for
the two groups, correlate with their different $\hat{\beta}$, as
expected from the idealized cyclone simulations (Fig.~\ref{fig:Theory fig}).
These make the case that indeed the cumulative $\beta$-drift, integrated
over all the cyclones in the group is the mechanism for the observed
westward drift of the cyclones. In this regard, we note that the westward
drift in itself is surprising, since the PCs, if possessing considerable
velocities as far equatorward as the CPCs, would advect the CPCs eastward
around both poles. But, seeing that the longitudinal displacements are
similar between the CPCs, the PCs and the CMs (lower panel of Fig.~\ref{fig: CM observations}f),
we conclude that advective steering of the CPCs by the PC winds is
unimportant, or is offset by the CM's westward drift. An important
intuition for the contrast between the drift rates of the north and
south poles can be derived from the fact that the ring of CPCs at
the north pole is much more concentric (to the pole) than at the south
pole (Fig.~\ref{fig: cart_trajectories}). Therefore, the
CM at the north is closer to the ``rest'' position at the pole,
leading to a smaller $\beta$ of the CM and a respectively smaller
westward drift, than at the south. Thus, if the cyclones were put
such that their CM rests exactly at the pole, no westward drift would
be expected.

Another piece of important information that is still missing about these polar
cyclones relates to the mechanism driving them and maintaining them
against dissipation. In a recent barotropic numerical study, a step was taken in that direction, showing that initial small-scale
turbulent conditions can ``cool down'' to form CPCs around a polar
cyclone \cite{siegelman2022polar}. This study is in agreement with
observational evidence showing that kinetic energy in the Jovian polar
region follows an inverse energy cascade, where energy originating
in small-scale turbulence is passed to the large-scale cyclones \cite{moriconi2020turbulence,siegelman2022moist}.
Regarding the depth of the cyclones, the $\beta$-drift framework,
explaining many observations, suggests that the cyclones are deep \cite{gavriel2022oscillatory}.
This is in agreement with the small Rossby number near the poles,
suggesting the 2D Taylor-Proudman theorem regime \cite{Busse1976,vallis2017atmospheric}.
However, there is still a gap towards understanding the mechanisms
driving the cyclones, which might reveal what determines their size,
strength and form. Finally, the unique conditions in Jupiter's poles,
provided striking opportunity to test, in the physical world, theories
derived for explaining the motion of tropical cyclones on Earth \cite{fiorino1989some,smith1990analytical,chan1995interaction},
where their transient nature and complex environment make the task
of diagnosing their motion extremely difficult. These theories prove
themselves consistently in Jupiter's polar cyclones, and assist in
constructing the physical picture behind their surprising existence.

\section*{Open Research}
No new data sets were generated during the current study. The data
analyzed in this study were published by Ref.\cite{mura2022five_years} (DOI: \url{https://doi.org/10.1029/2022JE007241}), as cited in the
text. The simulations were run with the Dedalus solver \cite{burns2020dedalus} (DOI: \url{https://doi.org/10.1103/PhysRevResearch.2.023068}).

\subsection*{acknowledgments}
This research has been supported by the Minerva Foundation with funding
from the Federal German Ministry for Education and Research and the
Helen Kimmel Center for Planetary Science at the Weizmann Institute
of Science. We thank Keren Duer, Eli Galanti, Or Hadas, Rei Chemke and Quentin Nicolas for insightful conversations and helpful feedback. \nocite{chan1987analytical}





%
\subsection*{Competing Interests}

Authors declare that they have no competing interests.

\bibliographystyle{naturemag}
\bibliography{Nimrodbib}

\renewcommand{\thefigure}{{\arabic{figure}}}
\setcounter{figure}{0}     
\renewcommand{\thetable}{S\arabic{table}} 
\renewcommand{\thefigure}{S\arabic{figure}}
\renewcommand{\theHtable}{Supplement.\thetable}
\renewcommand{\theHfigure}{Supplement.\thefigure}
\appendix
\onecolumn

\part*{Supplementary Information}

\noindent \begin{flushleft}
\textbf{Movie S1.} (\href{https://agupubs.onlinelibrary.wiley.com/action/downloadSupplement?doi=10.1029%2F2023GL103635&file=2023GL103635-sup-0002-Movie+SI-S01.mp4}{Link}) A movie showing the development of the $\beta$-gyres.
This is the full time evolution, from which the three snapshots of
Fig.~2 in the main text are taken. The left panel describes the vorticity
and the right panel is the stream-function. All variables are scales
according to SI section 2.
\par\end{flushleft}

\noindent \begin{flushleft}
\textbf{Movie S2. }(\href{https://agupubs.onlinelibrary.wiley.com/action/downloadSupplement?doi=10.1029%2F2023GL103635&file=2023GL103635-sup-0003-Movie+SI-S02.mp4}{Link}) A movie showing the vorticity evolution of the
6 identical cyclones $\beta$-plane experiment. The model is described
in SI section 3.
\par\end{flushleft}

\noindent \begin{flushleft}
\textbf{Movie S3. }(\href{https://agupubs.onlinelibrary.wiley.com/action/downloadSupplement?doi=10.1029%2F2023GL103635&file=2023GL103635-sup-0004-Movie+SI-S03.mp4}{Link}) A movie showing the vorticity evolution of the
equivalent cyclone $\beta$-plane experiment. The model is described
in SI section 3.
\par\end{flushleft}

\section{Cyclone paths from the observations}

Here we prescribe the methods we used for plotting the trajectories
in Fig.~1. For this study we used the data acquired by the JIRAM
instrument, as published in Ref.\cite{mura2022five_years}, where the
locations of the centers of the polar cyclones in the north and south
poles are laid out for Juno perijoves (PJs), where they were located
in terms of longitude ($^{\circ}$W) and latitude ($^{\circ}$N).
The first step in the analysis was to convert these locations to a
Cartesian grid according to 
\begin{equation}
x=-R_{{\rm J}}\cos\theta\,\sin\lambda,\;\;\;y=R_{{\rm J}}\cos\theta\,\cos\lambda,\label{eq: Polar to Cart}
\end{equation}
where $R_{{\rm J}}$ is the radius of Jupiter, $\theta$ is latitude
and $\lambda$ is longitude. The planetary radius here, calculated
as $R_{{\rm J}}=\left(\nicefrac{R_{{\rm e}}^{2}}{R_{{\rm p}}}\right)=76,452$
km \citep{hyder2022exploring}, is the osculating radius, taking Jupiter's
oblateness into account. This data, per PJ, is temporally interpolated
separately for the $x$ and $y$ coordinates. The blue arrows of Fig.~1e-f
are moving averages over the interpolated data, with an averaging
window of $\sim1.3$ years (endpoints are disposed; i.e., data points
$0.65$ years from the start or the end are not shown in the moving
average). 

\section{Beta-drift simulations}

\begin{figure}
\begin{centering}
\includegraphics[width=1\columnwidth]{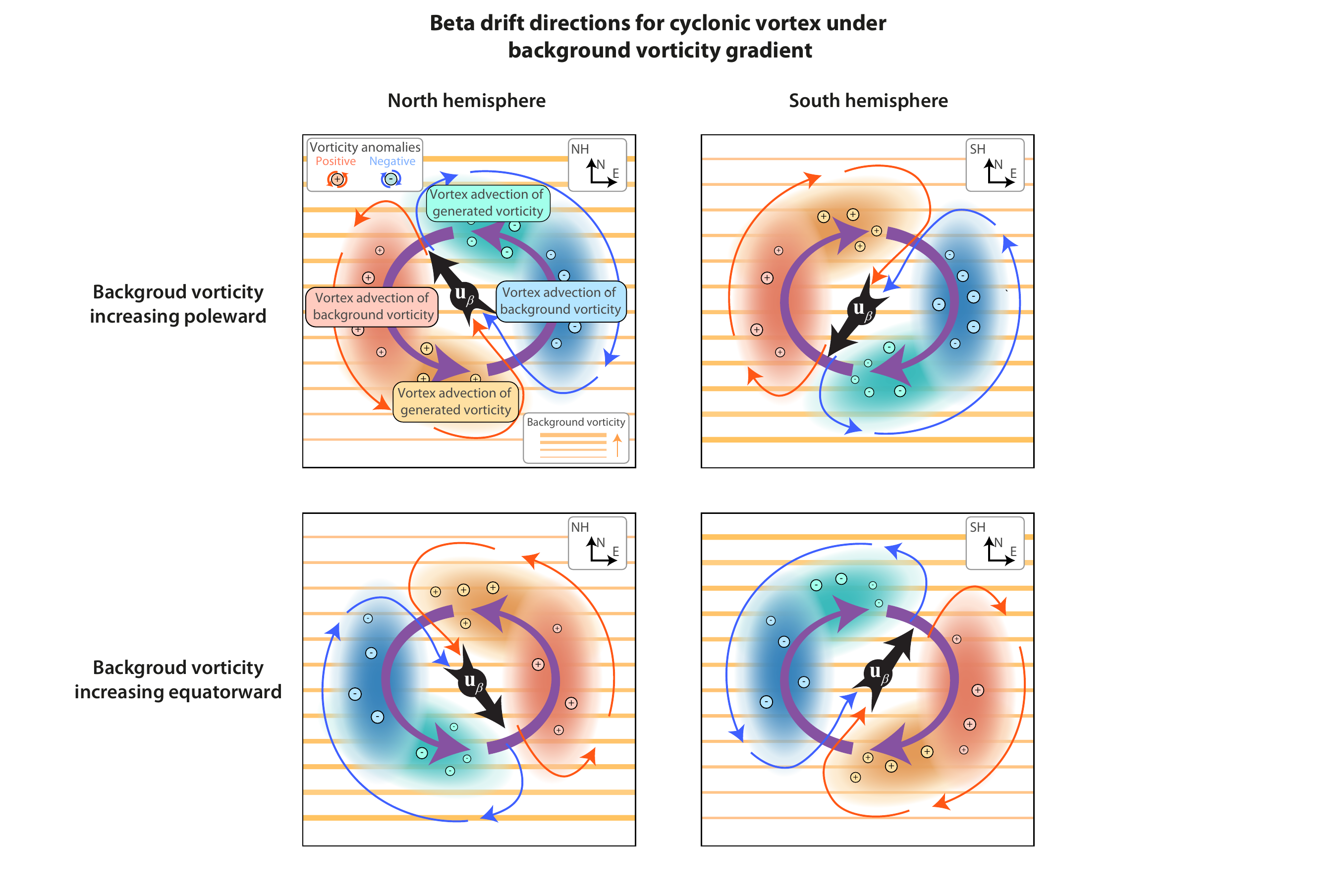}
\par\end{centering}
\caption{Generalization of Fig.~2e of the main text. Here the upper left panel
is the same as Fig.~2e. The left column represents $\beta$-drift
of a cyclonic vortex in the north hemisphere and the right column
represents it for the southern hemisphere. In the upper (lower) row,
background vorticity is increasing poleward (equatorward).}
\end{figure}
In this section, the model used for producing Fig.~2 is described. 

\subsection{Model equations}

We begin with the 2D barotropic vorticity equation, 
\begin{equation}
\frac{D}{Dt}\left(\xi+f\right)=0,\label{eq: Barotropic PV equation}
\end{equation}
where
\begin{equation}
\frac{D}{Dt}=\left(\frac{\partial}{\partial t}+\mathbf{u}\cdot\nabla\right)
\end{equation}
is the material derivative, $\mathbf{u}$ is the velocity vector,
$\xi=\nabla\times\mathbf{u}$ is relative vorticity and $f$ is the
vertical component of the planetary vorticity, or, in the context
of this study, any background vorticity profile. Eq.~\ref{eq: Barotropic PV equation}
is then partitioned, as suggested by Ref.{smith1990numerical},
into two coupled equations. One equation describes the advection of
a vortex by $\beta$-drift as
\begin{equation}
\frac{\partial\xi_{{\rm v}}}{\partial t}+\mathbf{u}_{\beta}\cdot\xi_{{\rm v}}=0,
\end{equation}
where $\xi_{{\rm v}}=\nabla\times\mathbf{u}_{{\rm v}}$ describes
the relative vorticity of a symmetric vortex, $\mathbf{u}_{{\rm v}}$
is the velocity vector of the vortex, and $\mathbf{u}_{\beta}$ is
the velocity vector of the $\beta$-drift. The second equation is
\begin{equation}
\frac{\partial\xi_{{\rm g}}}{\partial t}+\left(\mathbf{u}_{{\rm v}}+\mathbf{u}_{{\rm g}}\right)\cdot\nabla\left(\xi_{{\rm g}}+f\right)+\left(\mathbf{u}_{{\rm g}}-\mathbf{u}_{\beta}\right)\cdot\nabla\xi_{{\rm v}}=0,\label{eq: Full cons equation for generated vort}
\end{equation}
where $\xi_{{\rm g}}=\nabla\times\mathbf{u}_{{\rm g}}$ describes
the ``generated'' relative vorticity, and $\mathbf{u}_{{\rm g}}$,
the generated velocity vector, describes the generation of vorticity
anomalies (relative to the symmetric vortex) by the advection of net
vorticity, in a grid moving with \textbf{$\mathbf{u}_{\beta}$}. The
term $\mathbf{u}_{{\rm v}}\cdot\nabla\xi_{{\rm v}}$ was eliminated
for Eq.~\ref{eq: Full cons equation for generated vort} as this
prescribed vortex flow is symmetric around the vortex and flows along
lines of equal $\xi_{{\rm v}}$. In this formulation, it is assumed
that $\mathbf{u}_{\beta}$ is well represented by $\mathbf{u}_{{\rm g}}\left(x=0,\:y=0\right)$,
where the coordinates $\left(x=0,\:y=0\right)$ represent the center
of the cyclone in the moving grid.

For the numerical simulation, we use the Dedalus code package \citep{burns2020dedalus}.
Eq.~\ref{eq: Full cons equation for generated vort} in a $\beta$-plane,
is scaled according to 
\begin{equation}
\begin{array}{c}
\left(x,y\right)=R\left(\hat{x},\hat{y}\right),\;\;\;t=\frac{R}{V}\hat{t},\\
\mathbf{u}=V\hat{\mathbf{u}},\;\;\;\xi=\frac{V}{R}\hat{\xi},\;\;\;f=f_{0}+\beta R\hat{y,}
\end{array}\label{eq: scales}
\end{equation}
where $V$ is the maximum velocity in the vortex, $R$ is the radius
of maximum velocity in the vortex, and $\beta$ is the linear slope
of the background vorticity. With this scaling, together with a numerical
viscosity term and a Newton relaxation sponge term ($F$), Eq.~\ref{eq: Full cons equation for generated vort}
becomes (removing the hat signs for clarity)
\begin{equation}
\frac{\partial\xi_{{\rm g}}}{\partial t}+\left(\mathbf{u}_{{\rm v}}+\mathbf{u}_{{\rm g}}\right)\cdot\nabla\left(\xi_{{\rm g}}+\hat{\beta}y\right)+\left(\mathbf{u}_{{\rm g}}-\mathbf{u}_{\beta}\right)\cdot\nabla\xi_{{\rm v}}=\frac{1}{{\rm Re}}\nabla^{2}\xi_{{\rm g}}+F,\label{eq: Full cons equation for generated vort scaled}
\end{equation}
where 
\begin{equation}
\hat{\beta}=\frac{R^{2}\beta}{V},\label{eq: beta hat}
\end{equation}
and ${\rm Re}$ is a Reynolds number determining the strength of viscosity.
The trap, $F$, is aimed to relax the relative vorticity to zero far
away from the vortex, so as to avoid numerical boundary effects. Here,
we define $F$ as 
\begin{equation}
F=-\frac{1}{2}\frac{\xi_{{\rm g}}}{\tau_{{\rm trap}}}\left(1-\tanh\left(A\left(r_{{\rm trap}}-r\right)\right)\right),
\end{equation}
where $\tau_{{\rm trap}}$ defines the relaxation time scale, $r=\frac{1}{R}\sqrt{x^{2}+y^{2}}$
is a scaled distance from the center of the cyclone, and $r_{{\rm trap}}$
is the value of $r$ from which the relaxation begins, using a step
function with a $\tanh$ form and a sharpness amplitude $A$. The
reasoning by which Eq.~1 in the main text represents the leading
order of Eq.~\ref{eq: Full cons equation for generated vort scaled}
comes from assuming $\mathbf{u}_{{\rm g}}\ll\mathbf{u}_{{\rm v}}$,
which is well supported in the model run (Fig.~2a-d), so that in
the domain (far from the trap effect), we have 
\begin{equation}
\frac{\partial\xi_{{\rm g}}}{\partial t}+\mathbf{u}_{{\rm v}}\cdot\nabla\left(\xi_{{\rm g}}+\hat{\beta}y\right)+\left(\mathbf{u}_{{\rm g}}-\mathbf{u}_{\beta}\right)\cdot\nabla\xi_{{\rm v}}\approx0,\label{eq: Full cons equation for generated vort scaled-1}
\end{equation}
where ${\rm Re^{-1}}\ll1$. The reason why $\mathbf{u}_{{\rm v}}\cdot\nabla\left(\xi_{{\rm g}}+\hat{\beta}y\right)\gg\left(\mathbf{u}_{{\rm g}}-\mathbf{u}_{\beta}\right)\cdot\nabla\xi_{{\rm v}}$
is harder to justify and is discussed in depth by Ref.\cite{smith1990analytical}.
Here, we suffice noting that at the beginning of the motion, $\left(\mathbf{u}_{{\rm g}}-\mathbf{u}_{\beta}\right)$
is almost zero, but also later, at the core area where $\nabla\xi_{{\rm v}}$
is significant, $\mathbf{u}_{{\rm g}}\approx\mathbf{u}_{\beta}$ so
that $\left(\mathbf{u}_{{\rm g}}-\mathbf{u}_{\beta}\right)$ is very
small.

\subsection{Model run definitions}

The actual equations supplied to the model solve one variable, $\psi_{{\rm g}}$,
the generated streamfunction, in the form 
\begin{equation}
\frac{\partial}{\partial t}\left(\nabla^{2}\psi_{{\rm g}}\right)+\left(\mathbf{u}_{{\rm v}}+\mathbf{u}_{{\rm g}}\right)\cdot\nabla\left(\nabla^{2}\psi_{{\rm g}}+\hat{\beta}y\right)+\left(\mathbf{u}_{{\rm g}}-\mathbf{u}_{\beta}\right)\cdot\nabla\xi_{{\rm v}}+\tau=\frac{1}{{\rm Re}}\nabla^{4}\psi_{{\rm g}}-\frac{1}{2}\frac{\nabla^{2}\psi_{{\rm g}}}{\tau_{{\rm trap}}}\left(1-A\tanh\left(r_{{\rm trap}}-r\right)\right),
\end{equation}
where the relation between vorticity and streamfunction, $\xi_{{\rm g}}=\nabla^{2}\psi_{{\rm g}}$,
is used. The velocities in the $\left(x,y\right)$ directions are
$\left(u_{{\rm g}},v_{{\rm g}}\right)=\left(-\frac{\partial\psi_{{\rm g}}}{\partial y},\frac{\partial\psi_{{\rm g}}}{\partial x}\right)$.
$\tau$ is a tau term required for solution in the spectral methods
used by Dedalus (see Ref.\cite{burns2020dedalus}), which also require
a gauge equation,
\begin{equation}
\iint\psi_{{\rm g}}dx\:dy=0,
\end{equation}
over the domain. The scaled prescribed vortex profile used for the
model (suggested by Ref.\cite{chan1987analytical}, with $b=1$) is
\begin{equation}
U_{{\rm v}}=r\exp\left(1-r\right),\label{eq: Tangential velocity profile}
\end{equation}
where $U_{{\rm v}}$ is a the tangential velocity of the vortex. This
profile translates into the $\left(x,y\right)$ velocity components
\begin{equation}
\begin{array}{cc}
u_{{\rm v}}= & -U_{{\rm v}}\sin\left(\tan^{-1}\left(\frac{y}{x}\right)\right),\,{\rm and}\\
v_{{\rm v}}= & U_{{\rm v}}\cos\left(\tan^{-1}\left(\frac{y}{x}\right)\right),
\end{array}\label{eq: Cartesian velocity profiles}
\end{equation}
respectively. Subsequently, the vortex vorticity, calculated as $\xi_{{\rm v}}=\frac{1}{r}\frac{\partial}{\partial r}\left(rU_{{\rm v}}\right)$,
is
\begin{equation}
\xi_{{\rm v}}=\left(2-r\right)\exp\left(1-r\right).\label{eq: xi_v scaled}
\end{equation}
The equations are solved with a Fourier spectral base, implying double
periodic boundary conditions. These conditions are not important in
the solution due to the sponge term, forcing the vorticity field to
become zero away from the center. 

\paragraph*{Model details for Fig.~2a-c}

In this run, we solve the equations described in this section with
a resolution of $1024{\rm \times1024}$, in a domain with unit-less
size $18\times{\rm 18}$. The numbers for this simulation are: Re$=1000$,
$r_{{\rm trap}}=16$, $\tau_{{\rm trap}}=0.1$ $A=20$.

\paragraph*{Model details for Fig.~2f}

Here, a set of simulations are conducted to test how changes in $\hat{\beta}$
lead to changes in the time evolution of $\mathbf{u}_{\beta}$. The
simulations run with a grid resolution of $512\times512$ and a unit-less
domain size of $40\times40$. The numbers for all of the set's simulations
are: Re$=400$, $r_{{\rm trap}}=38$, $\tau_{{\rm trap}}=0.1$ $A=20$.
The $\hat{\beta}$ values for the different runs are $\left\{ 0.0001,0.001,0.003,0.005,0.01,0.05,0.1\right\} $.

\section{Derivation of an idealized \textquotedblleft center of mass\textquotedblright{}
for shielded cyclones}

\begin{figure}
\begin{centering}
\includegraphics[width=0.8\columnwidth]{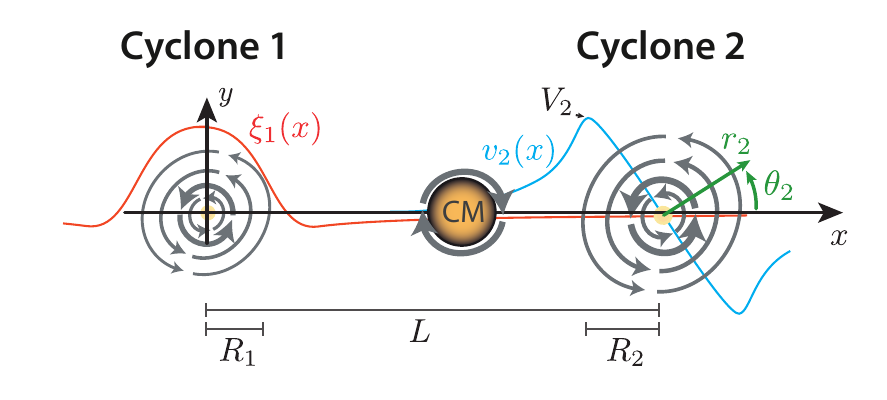}
\par\end{centering}
\caption{An illustration of the geometry for deriving a \textquotedblleft center
of mass\textquotedblright{} between 2 cyclones.\label{fig: Mass_derivation_geo}}
\end{figure}
In this section, a derivation is performed for finding the analog
of ``mass'' for a cyclone in a multiple cyclones system. We only
consider here shielded cyclones, i.e., cyclones with a ring of anticyclonic
vorticity around them. In this shielded case, the interactions between
cyclones are determined by the effect of the vorticity gradient of
one cyclone on another \citep{gavriel2021number,gavriel2022oscillatory},
rather than the advective interactions of one cyclone's tangential
velocity on another. Indeed, the polar cyclones of Jupiter are suspected
to be shielded based on observations \citep{gavriel2021number} and
on stability in shallow-water model simulations \citep{li2020}.

For this derivation, we consider the interaction between two interacting
cyclones (Fig.~\ref{fig: Mass_derivation_geo}). The instantaneous
vorticity gradient force (put here in terms of averaged acceleration)
of Cyclone 1 on Cyclone 2 in the $x$ direction can be shown \citep{gavriel2021number}
to be proportional to 
\begin{equation}
\pi R_{2}^{2}\frac{\partial\overline{u_{\beta,2}}}{\partial t}\approx\iint\xi_{1}v_{2}dS,\label{eq: cyc2 force integral}
\end{equation}
where $R_{i}$ is the cyclone's radius of maximum velocity, $\overline{u_{\beta,i}}$
is the mean, over the cyclone's core, drift velocity (by the vorticity
gradient forces) in the $x$ direction, $\xi_{i}$ is the vorticity
profile of a cyclone, $v_{i}$ is the $y$ component of a cyclone's
tangential velocity, $dS$ is the integration area over the cyclone's
core, and $i$ is the index of the cyclone ($1$ or $2$ for this
derivation). The dimensional forms of Eqs.~\ref{eq: Tangential velocity profile}-\ref{eq: xi_v scaled}
for a cyclone's velocity and vorticity profiles are 
\begin{equation}
v_{i}=V_{i}\frac{r_{i}}{R_{i}}\exp\left(1-\frac{r_{i}}{R_{i}}\right)\cos\left(\theta_{i}\right),
\end{equation}
\begin{equation}
\xi_{i}=\frac{V_{i}}{R_{i}}\left(2-\frac{r_{i}}{R_{i}}\right)\exp\left(1-\frac{r_{i}}{R_{i}}\right),
\end{equation}
where $r_{i}$ and $\theta_{i}$ represent the polar coordinates relative
to cyclone $i$, and $V_{i}$ is the maximum velocity of the cyclone.
A series expansion of $\xi_{1}$ around the center of Cyclone 2 ($r_{1}\rightarrow\sqrt{\left(L+r_{2}\cos\theta_{2}\right)^{2}+\left(r_{2}\sin\theta_{2}\right)^{2}}$,
where $L$ is the distance between the cyclones) gives
\begin{equation}
\xi_{1}\approx\frac{V_{1}}{R_{1}^{3}}\exp\left(1-\frac{L}{R_{1}}\right)\left(R_{1}\left(2R_{1}-L\right)+r_{2}\cos\theta_{2}\left(L-3R_{1}\right)\right),
\end{equation}
or, assuming $L\gg R_{1}$, 
\begin{equation}
\xi_{1}\approx\frac{V_{1}L}{R_{1}^{3}}\exp\left(1-\frac{L}{R_{1}}\right)\left(r_{2}\cos\theta_{2}-R_{1}\right).
\end{equation}
Similarly expanding $v_{2}$ around $r_{2}\rightarrow0$, we get 
\begin{equation}
v_{2}=V_{2}\frac{r_{2}}{R_{2}}e\cos\left(\theta_{2}\right).
\end{equation}
Plugging the expansions in Eq.~\ref{eq: cyc2 force integral}, after
preforming the integration ($dS=r_{2}dr_{2}d\theta_{2}$ in the boundaries
$\left\{ 0\leq r_{2}\leq R_{2}\right\} $ and $\left\{ 0\leq\theta_{2}\leq2\pi\right\} $)
gives
\begin{equation}
\frac{\partial\overline{u_{\beta,2}}}{\partial t}\approx\exp\left(2-\frac{L}{R_{1}}\right)\frac{LR_{2}V_{1}V_{2}}{4R_{1}^{3}}.\label{eq: F1 on 2}
\end{equation}
Similarly, for Cyclone 1
\begin{equation}
\frac{\partial\overline{u_{\beta,1}}}{\partial t}\approx-\exp\left(2-\frac{L}{R_{2}}\right)\frac{LR_{1}V_{1}V_{2}}{4R_{2}^{3}}.\label{eq: F2 on 1}
\end{equation}
Multiplying Eq.~\ref{eq: F1 on 2} by $\frac{1}{R_{2}^{4}}\exp\left(-\frac{L}{R_{2}}\right)$
and Eq.~\ref{eq: F2 on 1} by $\frac{1}{R_{1}^{4}}\exp\left(-\frac{L}{R_{1}}\right)$we
get 
\begin{equation}
\exp\left(-\frac{L}{R_{2}}\right)\frac{1}{R_{2}^{4}}\frac{\partial\overline{u_{\beta,2}}}{\partial t}\approx\exp\left(2-\frac{L}{R_{1}}-\frac{L}{R_{2}}\right)\frac{LV_{1}V_{2}}{4R_{1}^{3}R_{2}^{3}},\label{eq: F1 on 2-1}
\end{equation}
\begin{equation}
\exp\left(-\frac{L}{R_{1}}\right)\frac{1}{R_{1}^{4}}\frac{\partial\overline{u_{\beta,1}}}{\partial t}\approx-\exp\left(2-\frac{L}{R_{1}}-\frac{L}{R_{2}}\right)\frac{LV_{1}V_{2}}{4R_{1}^{3}R_{2}^{3}}.\label{eq: F2 on 1-1}
\end{equation}
Thus, summing Eqs.~\ref{eq: F1 on 2-1} and \ref{eq: F2 on 1-1}
gives
\begin{equation}
\exp\left(-\frac{L}{R_{2}}\right)\frac{1}{R_{2}^{4}}\frac{\partial\overline{u_{\beta,2}}}{\partial t}+\exp\left(-\frac{L}{R_{1}}\right)\frac{1}{R_{1}^{4}}\frac{\partial\overline{u_{\beta,1}}}{\partial t}=0,\label{eq: F1 on 2-1-1}
\end{equation}
representing the ``center of mass'' between the cyclones that does
not feel any acceleration in this simplified case. Finally, we propose
that for a group of cyclones, the ``center of mass'' analogy of
``mass'' gets the form 
\begin{equation}
m_{i}=\frac{\exp\left(-\frac{L}{R_{i}}\right)}{R_{i}^{4}},\label{eq: Mass of 1 cyclone}
\end{equation}
where $L$ represents an estimate for the average distance between
cyclones in the group.

\section{Simulations for the motion of a group of cyclones compared to one
equivalent cyclone}

In this section the model formulation resulting in Fig.~3 is presented.
These simulations were also conducted with the Dedalus package \citep{burns2020dedalus}.
For these results we are simply solving the barotropic vorticity equation
(Eq.~\ref{eq: Barotropic PV equation}) scaled according to Eq.~\ref{eq: scales},
leading to 
\begin{equation}
\frac{\partial}{\partial t}\left(\nabla^{2}\psi\right)+\mathbf{u}\cdot\nabla\left(\nabla^{2}\psi+\frac{1}{2}\hat{\beta}y\left(1-\tanh\left(A\left(r_{{\rm trap}}-r\right)\right)\right)\right)+\tau=\frac{1}{{\rm Re}}\nabla^{4}\psi.
\end{equation}
Here, the $f$ term is multiplied by a trap function such that a discontinuity
in $f$ traps vorticity anomalies (similar to Ref.\cite{siegelman2022polar}).
In this model we conduct two simulations. In the first run, a group
of 6 cyclones are inserted in the initial conditions of $\psi$ according
to
\begin{equation}
\xi_{i}=\left(2-r_{i}^{b}\right)\exp\left[\frac{1}{b}\left(1-r_{i}^{b}\right)\right],\label{eq: xi_v scaled-1}
\end{equation}
where $i$ represents the index of the cyclone (from 1 to 6), $r_{i}$
is defined by 
\begin{equation}
r_{i}=\sqrt{\left(x-x_{i}\right)^{2}+\left(y-y_{i}\right)^{2}},\label{eq: xi_v scaled-1-1}
\end{equation}
 and $(x_{i},y_{i})$ are the initial coordinates of the center of
cyclone $i$. The initial (dimensional) locations of the 6 cyclones
were $\{\left(19.998,-14.000\right)$, $\left(15.854,-8.295\right)$,$\left(9.147,-10.474\right)$,$\left(9.147,-17.526\right)$,$\left(15.854,-19.705\right)$,$\left(14.000,-14.000\right)$$\}\times10^{3}{\rm \;km}$.
For Fig.~3a-c, the cyclones all have $R=1000\;{\rm km}$, $V=80$
ms$^{-1}$ and $b=0.78$. In the simulation for the equivalent cyclone,
only one such cyclone was planted in the initial conditions at the
location $\left(14.000,-14.000\right)\times10^{3}{\rm \;km}$. The
initial condition for $\psi$ ($\psi_{0}$) was solved from the Poisson
equation, 
\begin{equation}
\nabla^{2}\psi_{0}=\xi_{0},
\end{equation}
where $\xi_{0}=\underset{i}{\sum}\xi_{i}$. These simulations had
a grid resolution of $1024{\rm \times1024}$ and a domain size of
$\left(37.36\times37.36\right)\times10^{3}{\rm \;km}$. The numbers
for these simulations are: Re$=8\times10^{9}$, $r_{{\rm trap}}=37.095\times10^{3}{\rm \;km}$,
$A=20$, $\hat{\beta}=7\times10^{-3}$. The trajectories of the cyclones
shown in Fig.~3 were produced by searching the local maxima of vorticity
in each vorticity snapshot, using the \textit{maximum\_filter} function
of the \textit{scipy }package in Python. As the initial cyclones are
identical, the CM trajectory is calculated simply as an average of
the $\left(x,y\right)$ coordinates of all the cyclones per snapshot.

\subsection{Simulations with size variance}

To more generally test the CM analogy, simulations with variance in
the cyclones' sizes are conducted. In one experiment (Fig.~3d), the
6-cyclone simulation has $R=1000\;{\rm km}$, $V=80$ ms$^{-1}$ and
$b=0.78$, except one cyclone that has $R=500\;{\rm km}$. In another
case (Fig.~3e), a second cyclone is also set to $R=500\;{\rm km}$,
and a third cyclone is set to $R=1200\;{\rm km}$. The initial locations
of the 6 cyclones are the same as in the 6 identical cyclones case.
The orange line, representing the CM motion is calculated according
to Eq.~\ref{eq: Mass of 1 cyclone}, where $L$ is estimated as $6\times10^{3}{\rm \;km}$.
For the equivalent cyclone run, the starting location is according
to the starting position of the CM in each case. The characteristics
of the equivalent cyclone are determined as a ``mass''-weighted
(using Eq.~\ref{eq: Mass of 1 cyclone}) average of the characteristics
of the 6 cyclones. There may be a more straightforward way to estimate
the characteristics of the equivalent cyclone to be derived in future
studies. In these simulation cases (Fig.~3d-e), the motion of each
cyclone is completely different than the motion in the results of
Fig.~3a, but the CM motion is still highly correlated with the motion
of the equivalent cyclone.

\section{Center of mass calculations from the observations}

\begin{figure}
\begin{centering}
\includegraphics[width=1\columnwidth]{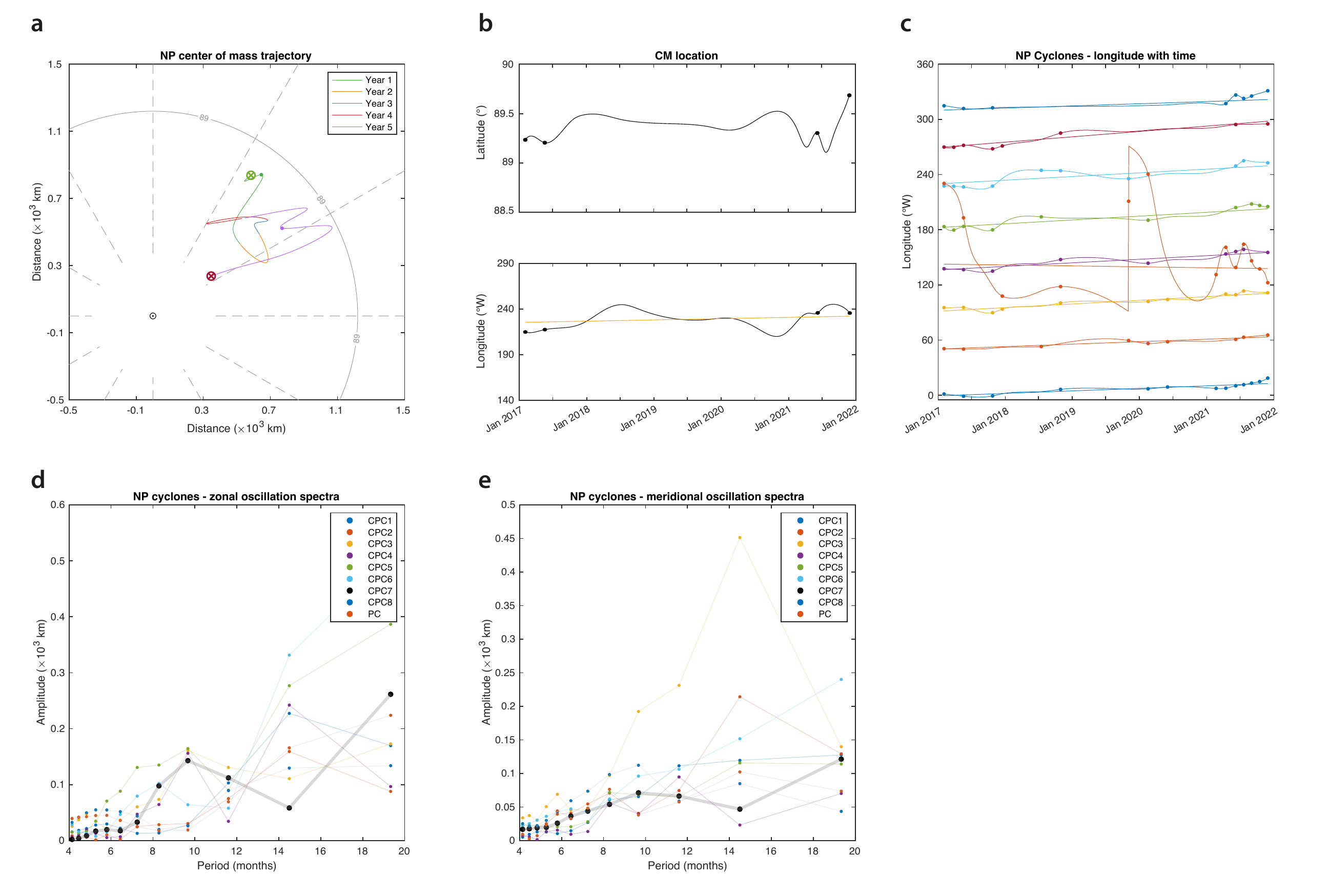}
\par\end{centering}
\caption{The same as Fig.~4 in the main text but for the north pole. As the
data is very infrequent, only the motion exhibited by the moving average
(Fig.~1f of the main text) has a significance for the analysis of
this study.\label{fig: north pole CM obs}}
\end{figure}
For Fig.~4a-e we only use data from the south pole, as the data from
the north pole is too infrequent for a meaningful analysis (See SI
Fig.~\ref{fig: north pole CM obs}). For Fig.~4c we translate back
the interpolated trajectories into polar coordinates according to
\begin{equation}
\theta=\cos^{-1}\left(\frac{\sqrt{x^{2}+y^{2}}}{R_{{\rm J}}}\right),\;\;\;\lambda=\tan^{-1}\left(\frac{y}{x}\right)-\frac{\pi}{2}.\label{eq: cart to polar}
\end{equation}
The straight lines in Fig.~4c are linear regression lines. To plot
the spectra (Fig.~4d-e) we calculated, for each cyclone, the cumulative
zonal ($l_{\lambda}$) and meridional ($l_{\theta}$) distances per
time step ($i$) according to 
\begin{equation}
\begin{array}{cc}
l_{\lambda}^{i}= & l_{\lambda}^{i-1}+R_{{\rm J}}\cos\left(\theta^{i}\right)\left(\lambda^{i}-\lambda^{i-1}\right),\\
l_{\theta}^{i}= & l_{\theta}^{i-1}+R_{{\rm J}}\cos\left(\theta^{i}-\theta^{i-1}\right),
\end{array}\label{eq: meridional and zonal time series-1}
\end{equation}
where $l_{\lambda}^{0}=l_{\theta}^{0}=0$. Then, after the mean and
linear trends are removed, a fast Fourier transform (FFT) was performed
on the time series defined by Eq.~\ref{eq: meridional and zonal time series-1}.
This procedure produces the amplitudes (in length units) per oscillation
period for each cyclone. Due to the long period of the collected data,
which is on the order of $50$ days ($\sim53$ days until PJ~34 and
$\sim43$ days after), only oscillation amplitudes with periods larger
than $4$ months are displayed. Also, we only show oscillations with
periods up to $20$ months, as oscillations with larger periods have
less than 3 repetitions in the total available data, containing $\sim5$
years.

For the calculation of the CM, the different cyclones are weighted
according to $m_{i}$ (Eq.~\ref{eq: Mass of 1 cyclone}), such that
the CM coordinates at each observation time are 
\begin{equation}
\left(x_{{\rm CM}},y_{{\rm CM}}\right)=\left(\frac{\sum x_{i}m_{i}}{\sum m_{i}},\frac{\sum y_{i}m_{i}}{\sum m_{i}}\right),
\end{equation}
where $i$ is the index of the cyclone ($1-9$ in the north pole and
$1-6$ in the south) and $L$ is assumed to be $6,000\;{\rm km}$,
in agreement with observational estimates (see the next section for
the sensitivity of the analysis to this choice). For $R_{i}$, we
use the data from Ref.\cite{adriani2020two}, which gives estimations
of the sizes for the cyclones in the north and south poles based on
Juno observations. As $R_{i}$ is the radius of maximum velocity,
we use the best fit from Ref.\cite{mura2022five_years} for the velocity
fields of the PCs during PJ4 \cite{grassi2018first} of $770$~km
and $970$~km for the north and south PCs, respectively. The respective
sizes of the PCs in Ref.\cite{adriani2020two} at PJ4, representing a
total size rather than size of maximum velocity, are $4440$~km and
$6930$~km for the north and south, respectively. Thus, to use the
data of Ref.\cite{adriani2020two}, the cyclone sizes are averaged over
PJs, and normalized by $\nicefrac{770}{4440}$ in the north and by
$\nicefrac{970}{6930}$ in the south. The resulting $R_{i}$ are presented
in Tab.~\ref{tab: Cyclones size}. For calculating $\hat{\beta}$
(Eq.~\ref{eq: beta hat}), the maximum velocities of the cyclones
are required, but there is no published data at the moment. We therefore
use the values estimated only for the PCs of $V_{i}=90$ and $95$
ms$^{-1}$ for the cyclones in the north and south poles, respectively
\cite{mura2022five_years}. Such variability, however, is less significant
for the determination of $\hat{\beta}$ relative to the squared $R_{i}$
variability.

Determining $\beta$ for the cyclones is more nuanced. In the $\beta$-plane
case (Fig.~3), $\beta$ is trivial since it is constant for all cyclones
at all times. In the polar case, however, different latitudes have
different values of $\beta$. In addition to that, while the direction
of $\beta$ is constant in the polar coordinates, the hydrodynamics
happen in the Cartesian coordinates, where $\beta$ changes direction
relative to the pole. Thus, to evaluate the drifts (i.e., evaluate
Eq.~\ref{eq: beta hat}) we set $\beta$ as equal between the cyclones
at each pole according to a temporal average of the vector sum of
the directional $\beta$ of each cyclone at each PJ for which we have
a size estimation. Specifically, for each PJ where we have location
and size data for the cyclones (only two such PJs in the north pole),
we calculate for each pole
\begin{equation}
\beta_{{\rm PJ\;}j}=\frac{1}{\sum_{{\rm cyc}\;i}m_{ij}}\left|\sum_{{\rm cyc}\;i}\overrightarrow{\beta_{ij}}m_{ij}\right|,
\end{equation}
which is a ``mass'' weighted vector sum of the directional beta
\begin{equation}
\overrightarrow{\beta_{ij}}=-\frac{2\Omega}{R_{{\rm J}}}\cos\left(\theta_{{\rm ij}}\right)\frac{\mathbf{r}_{ij}}{\left|\mathbf{r}_{ij}\right|},
\end{equation}
that has the magnitude of $\beta$ and a direction pointing towards
the pole ($\theta_{ij}$ and $\mathbf{r}_{ij}$ are the latitude and
location vector relative to the pole of cyclone $i$ at PJ $j$, respectively).
Jupiter's rotation rate is set as $\Omega=1.7585\times10^{-4}\;{\rm s}^{-1}$.
Lastly, the value we use for $\beta$ for each pole is set as
\begin{equation}
\beta=\frac{1}{n}\sum\beta_{{\rm PJ\;}j}
\end{equation}
over the $n$ PJ with sufficient data. This calculation leads to $\beta=\left\{ 4.862,\;8.896\right\} \times10^{-14}\left({\rm ms}\right)^{-1}$
for the north and south poles, respectively. The CM value of $\hat{\beta}$
for each pole is calculated by 
\begin{equation}
\hat{\beta}_{{\rm CM}}=\frac{\sum R_{i}^{2}m_{i}}{\sum m_{i}}\frac{\beta}{V},
\end{equation}
which gives the values $\hat{\beta}_{{\rm CM}}=\left\{ 3.771,\;7.257\right\} \times10^{-4}$
for the north and south poles, respectively.

The zonal velocities of the individual cyclones are calculated in
units of longitude ($^{\circ}$) per year and in metric units. For
the longitude case, the velocities were calculated from a linear fit
over the observed longitudes (Fig.~4c). For the metric case, first
the zonal cumulative distance ($l_{\lambda}$) was calculated for
each cyclone track according to Eq.~\ref{eq: meridional and zonal time series-1},
on which a linear fit was conducted for the mean velocity. The results
of both forms are laid out in Tab.~\ref{tab: Cyclones velocities}.
As discussed in the main text, two forms are presented for the zonal
velocity of the CM. For CM1 (Fig.~4f), the velocity is calculated
as the weighted average of $u_{i}$ (the zonal velocities of the individual
cyclones, Tab.~\ref{tab: Cyclones velocities}) according to 
\begin{equation}
u_{{\rm CM}1}=\frac{\sum u_{i}m_{i}}{\sum m_{i}}.
\end{equation}
For CM2, the velocities are calculated from the trajectory of the
CM by Eq.~\ref{eq: meridional and zonal time series-1}, where the
latitude and longitude are those of the CM and $u_{{\rm CM2}}$ results
from finally doing a linear fit on $l_{\lambda}^{{\rm CM}}$ for the
upper panel of Fig.~4f. For the lower panel of Fig.~4f, the velocities
result from a linear fit of the CM's longitude time series (lower
panel of Fig.~4b).

\begin{table}
\begin{centering}
{\small{}}%
\begin{tabular}{|c|c|c|c|c|c|c|c|c|c|c|}
\hline 
 & {\small{}PC} & {\small{}CPC1} & {\small{}CPC2} & {\small{}CPC3} & {\small{}CPC4} & {\small{}CPC5} & {\small{}CPC6} & {\small{}CPC7} & {\small{}CPC8} & {\small{}Mean}\tabularnewline
\hline 
\hline 
{\small{}North pole $R_{i}$ (km)} & {\small{}$969$} & {\small{}$789$} & {\small{}$757$} & {\small{}$771$} & {\small{}$732$} & {\small{}$891$} & {\small{}$823$} & {\small{}$810$} & {\small{}$756$} & {\small{}$811$}\tabularnewline
\hline 
{\small{}North pole $\nicefrac{m_{i}}{\sum m_{i}}$} & {\small{}$18.26\%$} & {\small{}$10.01\%$} & {\small{}$8.65\%$} & {\small{}$9.26\%$} & {\small{}$7.54\%$} & {\small{}$14.83\%$} & {\small{}$11.67\%$} & {\small{}$11.08\%$} & {\small{}$8.61\%$} & \tabularnewline
\hline 
{\small{}South pole $R_{i}$ (km)} & {\small{}$875$} & {\small{}$847$} & {\small{}$963$} & {\small{}$927$} & {\small{}$769$} & {\small{}$782$} & {\small{}-} & {\small{}-} & {\small{}-} & {\small{}$861$}\tabularnewline
\hline 
{\small{}South pole $\nicefrac{m_{i}}{\sum m_{i}}$} & {\small{}$17.57\%$} & {\small{}$15.95\%$} & {\small{}$22.40\%$} & {\small{}$20.48\%$} & {\small{}$11.42\%$} & {\small{}$12.18\%$} & {\small{}-} & {\small{}-} & {\small{}-} & \tabularnewline
\hline 
\end{tabular}{\small\par}
\par\end{centering}
\caption{An estimation for the cyclone sizes, and their relative weight, according
to the data from Ref.\cite{adriani2020two}.\label{tab: Cyclones size}}
\end{table}
\begin{table}
\begin{centering}
{\small{}}%
\begin{tabular}{|c|c|c|c|c|c|c|c|c|c|c|c|}
\hline 
 & {\small{}PC} & {\small{}CPC1} & {\small{}CPC2} & {\small{}CPC3} & {\small{}CPC4} & {\small{}CPC5} & {\small{}CPC6} & {\small{}CPC7} & {\small{}CPC8} & {\small{}CM1} & {\small{}CM2}\tabularnewline
\hline 
\hline 
{\small{}North pole $u_{i}$ (cm~s$^{-1}$)} & {\small{}$-1.10$} & {\small{}$1.52$} & {\small{}$1.42$} & {\small{}$2.22$} & {\small{}$2.24$} & {\small{}$2.51$} & {\small{}$2.19$} & {\small{}$3.49$} & {\small{}$0.99$} & {\small{}$1.55$} & {\small{}$0.23$}\tabularnewline
\hline 
{\small{}North pole $u_{i}$ ($\nicefrac{^{\circ}}{{\rm year}}$)} & {\small{}$-6.71$} & {\small{}$2.89$} & {\small{}$2.85$} & {\small{}$4.07$} & {\small{}$4.28$} & {\small{}$4.96$} & {\small{}$5.00$} & {\small{}$5.70$} & {\small{}$2.78$} & {\small{}$2.20$} & {\small{}$4.04$}\tabularnewline
\hline 
{\small{}South pole $u_{i}$ (cm~s$^{-1}$)} & {\small{}$2.95$} & {\small{}$4.39$} & {\small{}$3.43$} & {\small{}$3.42$} & {\small{}$4.37$} & {\small{}$4.39$} & {\small{}-} & {\small{}-} & {\small{}-} & {\small{}$3.72$} & {\small{}$3.47$}\tabularnewline
\hline 
{\small{}South pole $u_{i}$ ($\nicefrac{^{\circ}}{{\rm year}}$)} & {\small{}$9.24$} & {\small{}$7.83$} & {\small{}$7.69$} & {\small{}$7.74$} & {\small{}$9.05$} & {\small{}$7.86$} & {\small{}-} & {\small{}-} & {\small{}-} & {\small{}$8.17$} & {\small{}$4.85$}\tabularnewline
\hline 
\end{tabular}{\small\par}
\par\end{centering}
\caption{An estimation for the cyclone westward-drift velocities.\label{tab: Cyclones velocities}}
\end{table}

\subsection{Sensitivity of the results to the cyclones' weights}

\begin{figure}
\begin{centering}
\includegraphics[width=1\columnwidth]{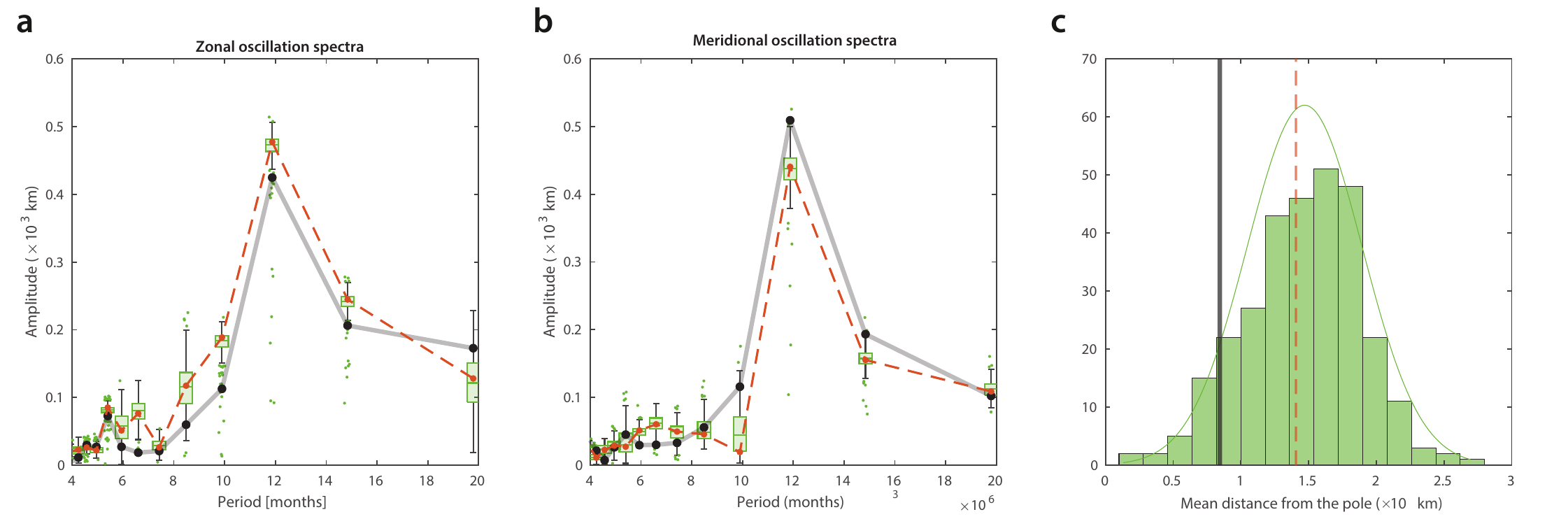}
\par\end{centering}
\caption{A sensitivity analysis of the CM in Jupiter's south pole to the cyclones'
parameters. (a and b) The oscillation spectra in the zonal (a) and
meridional (b) directions of the CM. The black curves represent a
CM trajectory weighted with the observational estimation of the cyclones'
sizes (Tab.~\ref{tab: Cyclones size}), which is the same as the
black curves in Fig.~4d-e. The orange dashed curves represent a CM
trajectory calculated with all cyclones weighted equally, such that
it represents a geometric mean. The green box plots are a statistical
representation of 300 randomly generated sets of weights (see text),
where the box is defined by the upper and lower quantiles (with the
median being the line inside the box), the black \textquotedblleft whiskers\textquotedblright{}
are the maximum and minimum values that are not outliers (where outliers
are values away from the box boundaries by 1.5 times the size of the
box), and the green dots are outliers. (c) A histogram showing the
amount of randomly generated sets (ordinate) that corresponds with
each range of calculated $R_{{\rm mean}}$ (see text). Here the green
curve represents a normal distribution of the random sets, the black
line is calculated using the observational estimated weights, and
the orange dashed line is calculated using equally weighted cyclones.\label{fig:Sensitivity_fig}}
\end{figure}
The following section serves to answer two questions: \textit{i})
what is the sensitivity of the analyses of Fig.~4 to the ``mass''
of each cyclone? \textit{ii}) How adequate is the estimation of $m_{i}$
(Eq.~\ref{eq: Mass of 1 cyclone}) determined in this study, when
viewed through the lens of the analyses of Fig.~4 of the main text?
To answer these, we compare the Tab.~\ref{tab: Cyclones size} values
against random sets of values and against a ``geometric mean'' set,
where each cyclone is weighted equally. Each random set is constructed
by assigning to each cyclone a random $R_{i}$ from a student's t-distribution
(Matlab's trnd() function) with the mean and variance of the third
row of Tab.~\ref{tab: Cyclones size}, and a sample size of 6. This
way the cyclones retain a distribution of sizes similar to the one
estimated from the observations. The weight of each cyclone is then
calculated, as before, by{\small{} $\nicefrac{m_{i}}{\sum m_{i}}$.}{\small\par}

For Fig.~\ref{fig:Sensitivity_fig}, 300 such random sets are generated.
In Fig.~\ref{fig:Sensitivity_fig}a-b the oscillation spectra (zonal
in a, and meridional in b) of the CM, calculated using these random
sets for the weights, are presented in the form of box plots (in green).
It can be seen that while the meridional spectra of the random sets
is relatively similar to the observational weighted set, in the zonal
spectra, using the observational values significantly reduces the
small period oscillations (periods of 6-10 months), and thus the observational
values better represent the ``real'' CM, which should ideally only
retain the one oscillation mode (\textasciitilde 12 months), forced
only by the planetary PV gradient. Another interesting aspect of Fig.~\ref{fig:Sensitivity_fig}a-b,
which supports our determination of the cyclones' weights, is that
for the orbital mode (\textasciitilde 12 months), the black curve
has a higher amplitude in the meridional direction and a lower amplitude
in the zonal direction than the majority of the random sets. This
signifies an elliptical orbit with a semi-major axis in the meridional
direction for the black case, similar to what we would expect from
an ideal representation of a CM for this motion, with a radial force
in the poleward direction (Fig.~3f). In Fig.~\ref{fig:Sensitivity_fig}c,
the mean distance of the CM's trajectory, 
\begin{equation}
R_{{\rm mean}}=\sqrt{\left(\frac{1}{n}\sum_{i=1}^{n}x_{{\rm CM},i}\right)^{2}+\left(\frac{1}{n}\sum_{i=1}^{n}y_{{\rm CM},i}\right)^{2}},
\end{equation}
where $n$ is the size of the interpolated data, is calculated for
each case. This parameter represents how close is the overall mean
position of the trajectory to the pole, and should be as close to
zero as possible for a ``physical'' representation of the CM (Fig.~3f).
It can be seen that indeed the black line, representing the observational
weights, has smaller $R_{{\rm mean}}$ than the majority of the random
sets and than the equally weighted set, supporting the use of these
observationally based weights to the CM analyses.

\paragraph{Sensitivity to the parameter $L$}

\begin{figure}
\begin{centering}
\includegraphics[width=1\columnwidth]{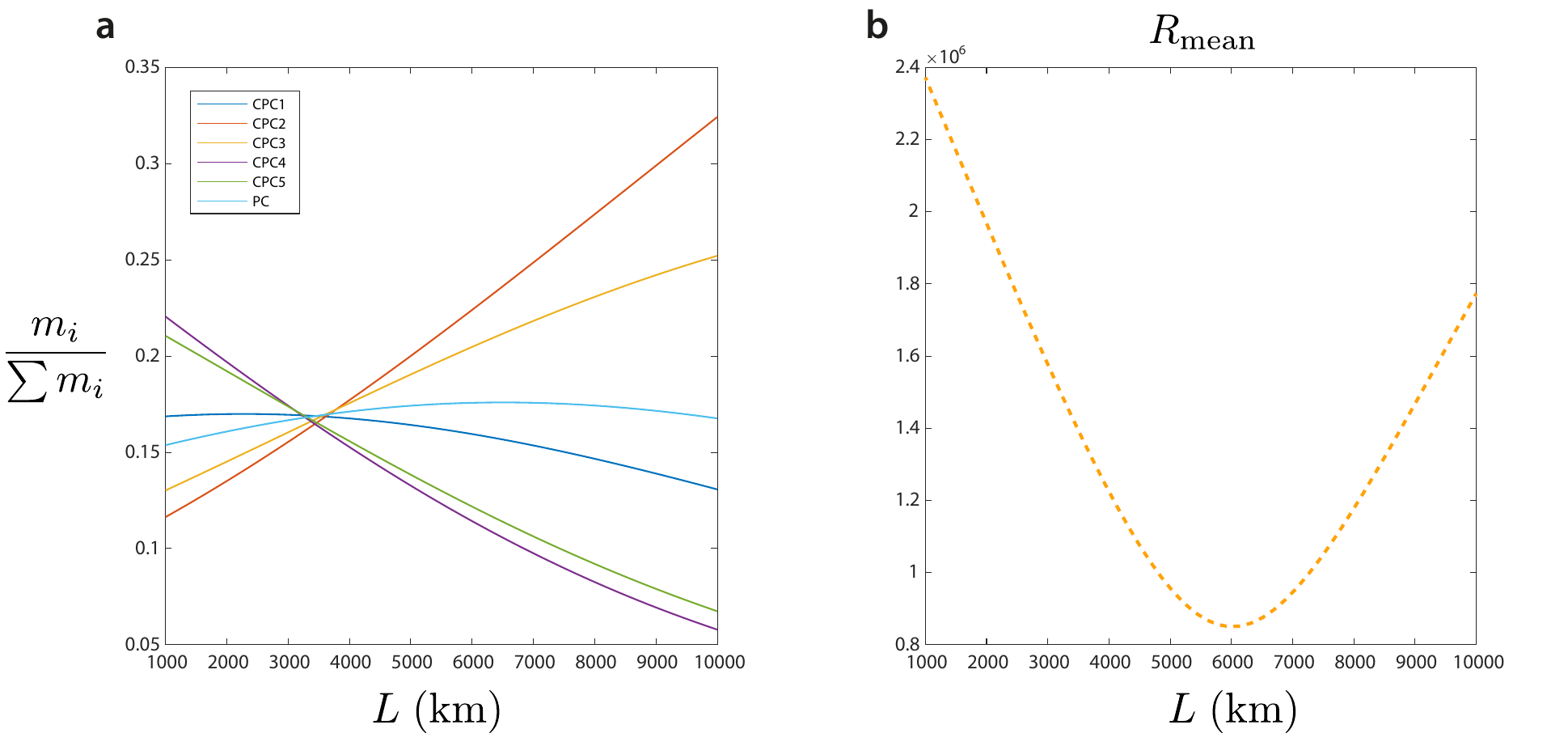}
\par\end{centering}
\caption{Sensitivity of $m_{i}$ (a) and the mean distance of the CM from the
pole (b) to the choice of $L$. This figure is only for the south
pole.\label{fig:Sensitivity_to_L}}
\end{figure}
Here, we see look at the sensitivity to the parameter $L$, which
represents a mean distance between the cyclones (Fig.~\ref{fig: Mass_derivation_geo}).
In Fig.~\ref{fig:Sensitivity_to_L}a, we see how the weights of each
cyclone in the south pole would change when changing $L$. Interestingly,
our choice ($L=6,000$ km, based on an estimation from Fig.~1 of
the main text) represents a minimum when considering how $R_{{\rm mean}}$
of the motion of the CM changes with $L$ (Fig.~\ref{fig:Sensitivity_to_L}b).
This choice of $L$, is also consistent with the minimum distance
between cyclones for stability ($L_{{\rm lim}}$) that was calculated
in Ref.\cite{gavriel2021number}, where it was calculated that $L_{{\rm lim}}\approx6R_{{\rm CPC}}$
and where the cyclone's radius of maximum velocity is $R_{{\rm CPC}}\approx1,000$
km.

\end{document}